\documentclass[twocolumn]{aastex701}
\usepackage[utf8]{inputenc}
\usepackage{textcomp}
\usepackage{tabularx}

\usepackage{placeins}
\usepackage{float}

\shorttitle{mid-infrared TDEs in AGN}
\shortauthors{Upadhyayula et al.}

\begin{document}

\title{Extreme AGN Variability in WISE:\\
Powerful Flares and Candidate Tidal Disruption Events in AGN}

\author[0009-0002-4104-8855]{Satya Teja Anuraag Upadhyayula}
\affiliation{University of Missouri -- Kansas City, Kansas City, MO 64110, USA}
\email{asubt2@umsystem.edu}

\author[0000-0003-2686-9241]{Daniel Stern}
\affiliation{Jet Propulsion Laboratory, California Institute of Technology, Pasadena, CA 91109, USA}
\email{daniel.k.stern@jpl.nasa.gov}

\author[0000-0002-8989-0542]{Kishalay De}
\affiliation{Department of Astronomy, 
Columbia University, 550 W 120th Street, New York, NY 10027, USA}
\affiliation{Center for Computational Astrophysics, Flatiron Institute, 162 5th Avenue, New York, NY 10010, USA}
\email{kd3038@columbia.edu}

\author[0009-0001-9034-6261]{Christos Panagiotou}
\affiliation{MIT Kavli Institute for Astrophysics and Space Research, Massachusetts Institute of Technology, Cambridge, MA 02139, USA}
\email{cpanag@mit.edu}

\author[0000-0002-3168-0139]{Matthew J. Graham}
\affiliation{Cahill Center for Astrophysics, California Institute of Technology, 1216 East California Boulevard, Pasadena, CA 91125, USA}
\email{mjg@caltech.edu}

\author[0000-0002-5956-851X]{K. E. Saavik Ford}
\affiliation{Department of Astrophysics, American Museum of Natural History, 200 Central Park West, New York, NY, 10024, USA}
\affiliation{Department of Science, CUNY Borough of Manhattan Community College, 199 Chambers Street, New York, NY 10007, USA}
\affiliation{Physics Program, CUNY Graduate Center, 365 5th Avenue, New York, NY 10016, USA}
\email{sford@amnh.org}

\author[0000-0002-9726-0508]{Barry McKernan}
\affiliation{Department of Astrophysics, American Museum of Natural History, 200 Central Park West, New York, NY, 10024, USA}
\affiliation{Department of Science, CUNY Borough of Manhattan Community College, 199 Chambers Street, New York, NY 10007, USA}
\affiliation{Physics Program, CUNY Graduate Center, 365 5th Avenue, New York, NY 10016, USA}
\email{bmckernan@amnh.org}

\author[0000-0002-8147-2602]{Murray Brightman}
\affiliation{Cahill Center for Astrophysics, California Institute of Technology, 1216 East California Boulevard, Pasadena, CA 91125, USA}
\email{mbright@caltech.edu}

\author[0000-0002-9654-1711]{J. A. Acevedo Barroso}
\affiliation{Jet Propulsion Laboratory, California Institute of Technology, Pasadena, CA 91109, USA}
\email{javier.a.acevedo.barroso@jpl.nasa.gov}

\author[0009-0000-1182-6420]{Daniel H. McIntosh}
\affiliation{University of Missouri -- Kansas City, Kansas City, MO 64110, USA}
\email{mcintoshdh@umkc.edu}

\begin{abstract}
We present first results from a systematic study of extreme mid-infrared variability of active galaxies across the $>10$-year baseline of the Wide-field Infrared Survey Explorer (WISE) mission.  WISE, which observed the full sky with a six-month cadence, is sensitive to transient events in both unobscured and obscured galaxies.  This paper focuses on extreme flares in active galactic nuclei (AGNs).  Previous studies identified flares such as tidal disruption events (TDEs) in inactive galaxies, typically from optical or X-ray surveys. However, flares are also expected in AGN and optical/X-ray searches are insensitive to events in dust-enshrouded AGN. In this pilot study, we present flares in two AGN identified from extreme mid-infrared color variability.  One flare was also detected at optical wavelengths.  While both events showed broadened H$\alpha$ emission several years after the flare peak, more recent spectroscopy reveals continued broadened emission only from the source with the optical flare.  After considering several potential physical causes of the flares, including supernovae and microlensing, we suggest both events are TDEs in AGN. Intriguingly, the optical flare decays much faster than the expected $t^{-5/3}$ fallback rate expected for TDEs in quiescent galaxies, suggesting either a partial disruption or the tidal disruption of a star by an intermediate mass black hole embedded in the accretion disk of an AGN.  This work demonstrates the power of WISE to identify the full census of TDEs, including events in dust-obscured AGN. Mid-infrared data probes the nature of the event and provides a bolometric measure of the flare energetics.
\end{abstract}


\section{Introduction}\label{sec:intro}

Tidal disruption events (TDEs) happen when a star comes too close to a supermassive black hole (SMBH) and is torn apart by tidal forces. This occurs when a star of mass $M_\star$ and radius $R_\star$ approaches a black hole of mass $M_{\rm BH}$ and passes within the tidal radius, $R_{\rm t} = R_\star (M_{\rm BH}/M_\star)^{1/3}$: this is the distance at which the gravitational pull across the star is stronger than the star's own self-gravity \citep{Hills1975, Rees1988}. After the star is disrupted, about half the debris stays bound to the black hole and falls back at a rate that drops as $t^{-5/3}$ for a typical TDE in a quiescent galaxy \citep{Rees1988, Phinney1989}. As the material returns, relativistic precession causes the streams to collide with each other, which helps the gas settle into an accretion disk \citep{Evans1989, Strubbe2009, Shiokawa2015}. This circularization of debris can produce a bright flare reaching luminosities of $10^{44}$--$10^{45}$\,erg\,s$^{-1}$ and lasting for months to years \citep{Lacy1982, Rees1988}.

TDEs were first expected to appear as bright, transient X-ray sources powered by accretion. However, observations over the past two decades have shown a more varied picture \citep[for a recent review, see][]{Gezari2021}. Some TDEs are X-ray bright, but many others shine mostly in the UV and optical with little or weak X-ray emission \citep{Auchettl2017}. This has led to different models. Some propose that TDE emission primarily comes from the newly formed accretion disk, while other models propose that shocks during debris circularization dominate the emission \citep{Guillochon2013, 2015ApJ...803...41M, Piran2015, Roth2016}.

The first TDEs were found as soft X-ray flares in ROSAT data \citep{Bade1996,Komossa1999}. Later, GALEX added UV detections \citep{Gezari2006, Gezari2008}. The discovery rate went up significantly with wide-field optical surveys like the Palomar Transient Factory \citep[PTF;][]{ptf}, the Panoramic Survey Telescope and Rapid Response System \citep[Pan-STARRS;][]{Chambers2016}, the All-Sky Automated Survey for Supernovae \citep[ASAS-SN;][]{Shappee2014}, the Asteroid Terrestrial-impact Last Alert System \citep[ATLAS;][]{Tonry2018}, and the Zwicky Transient Facility \citep[ZTF;][]{Bellm2019}. These surveys found that optical TDEs tend to occur in a specific type of host galaxy: quiescent Balmer-strong galaxies (also called E+A or post-starburst galaxies) and galaxies in the green valley, which lies between blue star-forming galaxies and red passive galaxies on the color vs mass diagram \citep{French2016, LawSmith2017, vanVelzen2021a}. This suggests that recent quenching of star formation or past galaxy mergers might boost TDE rates.

However, optical surveys miss TDEs hidden behind dusty regions or in dusty galaxies. If a TDE happens in a galaxy with a significant amount of obscuration along the line of sight, the soft X-ray, UV and optical light will be absorbed and the event will be missed by searches in those wavebands \citep{Roth2021}. This is where mid-infrared observations help. Dust that absorbs UV/optical light heats up and re-emits that energy in the infrared, creating what is called a ``dust echo'' \citep{Lu2016,vanVelzen2016}. Mid-infrared emission is much less affected by dust, so surveys at these wavelengths will find TDEs that higher energy surveys miss.

The Wide-field Infrared Survey Explorer (\textit{WISE}; \citealt{Wright2010}) and its NEOWISE reactivation \citep{Mainzer2011, 2014ApJ...792...30M} have imaged the full sky at $3.4$ and $4.6\,\mu$m since 2010, and the accumulated archive has been reorganized into time-domain catalogs and searched for variability with a range of techniques \citep[e.g.,][]{Meisner2023, Paz2024, Paz2026}. Early on, several groups showed that WISE could identify luminous AGN based on their red mid-infrared emission, which comes from hot dust near the central black hole \citep{Jarrett2011, Stern2012, Assef2013}. Building on this, \citet{Assef2018} created the ``WISE AGN Catalog,'' which contains 4.54 million AGN candidates selected with 90\% reliability (the R90 sample) across approximately 30,900\,deg$^2$ of extragalactic sky, and 20.9 million candidates with 75\% completeness (the C75 sample) over the same area. The R90 sample uses strict color cuts to keep contamination low. The six-month cadence and full-sky coverage of WISE make it well-suited for detecting slow transients and has a sufficient baseline ($\sim$ 15 years) to capture entire flares, as well as rare events like changing-look quasars. One recent study making use of this capability is \citet{masterson2024newpopulationmidinfraredselectedtidal}, which found 18 mid-infrared TDE candidates in nearby galaxies. They identified sources with strong W2 variability and used strict pre-flare color and brightness cuts to avoid AGN contamination. In parallel, the FLAIRES project \citep{Necker_2025} searched over 40 million galaxies for mid-infrared transients. Using a Bayesian Blocks algorithm to find flares, they identified 823 dust-echo candidates across the sky. Together, these studies show that mid-infrared surveys can find TDEs that are missed by optical surveys.

In this paper, we present results from our own search for extreme mid-infrared variability in WISE-selected AGN. While the studies mentioned above primarily focused on quiescent host galaxies, our work specifically targets the WISE AGN population, searching for TDE and flare signatures superimposed on pre-existing AGN activity. Section~\ref{sec:data} describes the WISE data and the parent AGN sample. Section~\ref{sec:methodology} explains the variability metrics we use to find flaring sources. Section~\ref{sec:results} presents two objects with extreme mid-infrared flares that we identify as TDE candidates, along with archival and follow-up spectroscopy. Section~\ref{sec:modeling} describes the dust echo modeling we use to measure the dust geometry and flare timing. Section~\ref{sec:discussion} explores different underlying physical scenarios that could cause a flare in the mid-infrared regime. Section~\ref{sec:conclusions} summarizes our findings.

Throughout this paper, we use Vega magnitudes for WISE mid-infrared photometry and AB magnitudes for optical photometry. Unless otherwise noted, we assume a flat $\Lambda$CDM cosmology with $H_0 = 70\, \mathrm{km\, s^{-1}\, Mpc^{-1}}$, $\Omega_m = 0.3$, and $\Omega_\Lambda = 0.7$.

\section{Data}
\label{sec:data}

\subsection{WISE and NEOWISE Observations}
\label{sec:wise}

WISE \citep{Wright2010} launched in December 2009 and imaged the full sky in four mid-infrared bands: W1 (3.4\,$\mu$m), W2 (4.6\,$\mu$m), W3 (12\,$\mu$m), and W4 (22\,$\mu$m). The cryogenic phase of the main mission ended in September 2010, after which the two longer wavelength bands (W3 and W4) ceased operation, though W1 and W2 continued to operate with no loss in sensitivity. The satellite was placed into hibernation in February 2011.

The spacecraft was reactivated in late 2013 as NEOWISE \citep{Mainzer2011, 2014ApJ...792...30M}. From then until August 2024, NEOWISE surveyed the sky in W1 and W2 with a roughly six-month cadence between epochs (epochs are averaged data over six-month intervals). During each pass, most sources were observed at least 12 times over a few days. The number of observations is higher near the ecliptic poles due to the spacecraft's orbit. In total, WISE and NEOWISE observed the full sky more than 24 times over nearly 15 years.  In this pilot study, we use the first 20 epochs of WISE/NEOWISE data, which corresponds to the available data at the start of this project.

\subsection{The WISE R90 AGN Catalog}
\label{sec:r90}

Our parent sample comes from the WISE AGN Catalog by \citet{Assef2018}, which identifies AGN candidates using mid-infrared colors 
\citep[e.g.,][]{Stern2012}. The catalog consists of two samples. The R90 sample is designed for high reliability: based on a detailed analysis of UV- to near-IR spectral energy distributions of $\sim 10^5$ sources in the 9 deg$^2$ Bo\"otes field, 90\% of sources that pass the R90 cuts should be real AGN. The C75 sample is designed for high completeness: based on the same Bo\"otes analysis, C75 captures 75\% of AGN, but comes with a higher contamination rate. For this pilot study, we use R90. 

The R90 selection uses a W1$-$W2 color cut that depends on W2 brightness. Fainter sources need redder colors to be included, which keeps reliability constant across all magnitudes. The catalog requires a point-source morphology in WISE, which has an angular resolution (full-width at half-maximum) of $6.1^{\prime \prime}$ in W1 and $6.4^{\prime \prime}$ in W2.  The catalog also requires no artifact flags and good detections in both W1 ($\geq 3 \sigma$) and W2 ($\geq 5\sigma$). Regions near the Galactic Plane and Galactic Center are excluded to avoid stars, as are areas around planetary nebulae, H\,{\sc ii} regions, and nearby galaxies. After these cuts, the R90 sample contains 4,543,530 AGN candidates across 30,093\,deg$^2$, corresponding to $\sim$151 AGN per square degree. This large sample provides an effective starting point for finding rare mid-infrared transients.

For this work, we obtained mid-infrared light curves by processing images from the WISE/NEOWISE surveys \citep{Wright2010, 2014ApJ...792...30M} released as part of the \texttt{unwise} project \citep{meisner2018}. We performed forced aperture photometry using a 2-pixel ($\approx 5.5$\arcsec) radius using a custom photometry pipeline (see \citet{de2023}, for details).

\section{Methodology}
\label{sec:methodology}

For this pilot study, to find AGN with extreme mid-infrared variability, 
we used W1$-$W2 color variability instead of flux variability since color changes are more sensitive to physical changes in the dust, namely large temperature shifts due to powerful nuclear flares or changes in the accretion state or accretion disk of the AGN. 


\subsection{Quality Filtering}
\label{sec:quality}

Before measuring variability, we applied quality cuts to remove spurious data from a pilot sample of 600,000 R90 AGN candidates. We required signal-to-noise of at least 10 in both W1 and W2 for each epoch, corresponding to photometric uncertainties $\lesssim$0.1 mag.  This decreased the sample to 214,750 objects. To avoid unrealistically small errors, we filtered out sources with less than 0.01 mag uncertainty, further reducing our pilot sample to 208,405 objects. 
We applied an additional requirement of W1$-$W2 $> 0$, which resulted in the loss of 14 objects. Based on a visual inspection, most of these appear to be bright ($r \leq 15$) Galactic stars with W1$-$W2 between $-0.2$ and zero (i.e., typical of stellar photospheres) that spuriously appear in the WISE AGN Catalog.  

After these cuts, we kept only objects with at least 15 epochs to ensure enough data for reliable variability measurements. This left us with a final sample of 99,634 high-reliability AGN candidates with robust, multi-epoch mid-infrared photometry. Future studies will explore more lenient quality cuts on a larger AGN sample.

\subsection{Variability Metrics}
\label{sec:metrics}

For each object, we calculated three metrics to quantify the variability in color $C =$ W1$-$W2.

\subsubsection{Weighted RMS}

The weighted root-mean-square (wRMS) measures the scatter in color, giving less weight to noisier points:
\begin{equation}
\mathrm{wRMS} = \sqrt{\frac{\sum_i w_i \,(C_i - \bar{C}_w)^2}{\sum_i w_i}},
\end{equation}
where $C_i$ is the color at epoch $i$, $\sigma_i$ is the corresponding uncertainty, $w_i = 1/\sigma_i^2$ is the weight, and $\bar{C}_w = \sum_i w_i C_i / \sum_i w_i$ is the weighted mean color. Larger values of wRMS indicate more scatter, implying stronger variability.

\subsubsection{Amplitude}

\begin{figure*}[t]
    \centering
    \includegraphics[width=0.85\linewidth]{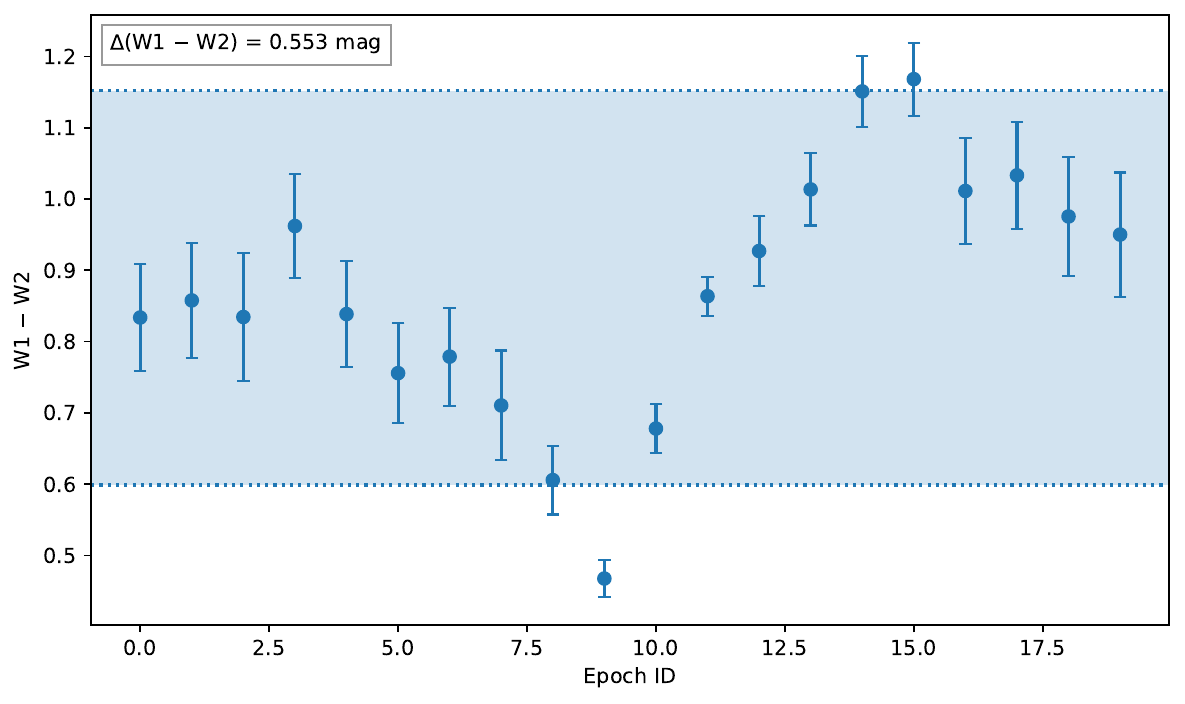}
    \caption {Illustration of the amplitude metric, showing the light curve of WISE J2150$-$1132. Photometry outside the 5th to 95th flux distribution are excluded to reduce contamination from potentially spurious light curve points.} 
    
    \label{fig:placeholder}
\end{figure*}

As a second metric, we consider the range over which the W1$-$W2 color varied across the WISE/NEOWISE mission.  In order to avoid possibly spurious outlier values, we define the amplitude $A$ as the difference between the 95th and 5th percentile values in the color range; i.e., for sources with 20 epochs of observations, we drop the bluest and reddest epoch color values.  For sources with fewer than 20 epochs, we compute the 95th and 5th percentile values via linear interpolation (calculating the mean) between the nearest ranked values. An example for this metric is shown in Figure~\ref{fig:placeholder}. 

\subsubsection{Reduced Chi-Squared}

Reduced chi-squared $\chi^2_\nu$ tests whether the color variations are larger than expected from measurement noise:
\begin{equation}
\chi^2_\nu = \frac{1}{N-1} \sum_i \left( \frac{C_i - \bar{C}_w}{\sigma_i} \right)^2,
\end{equation}
where $N$ is the number of epochs. For a source with constant color and Gaussian errors, the reduced chi-square should be approximately one. Values much larger than one indicate that the color is intrinsically variable \citep{Kozlowski2016}.

\subsection{Ranking and Selection}
\label{sec:ranking}

\begin{figure*}[t]
    \centering
    \includegraphics[width=0.85\textwidth]{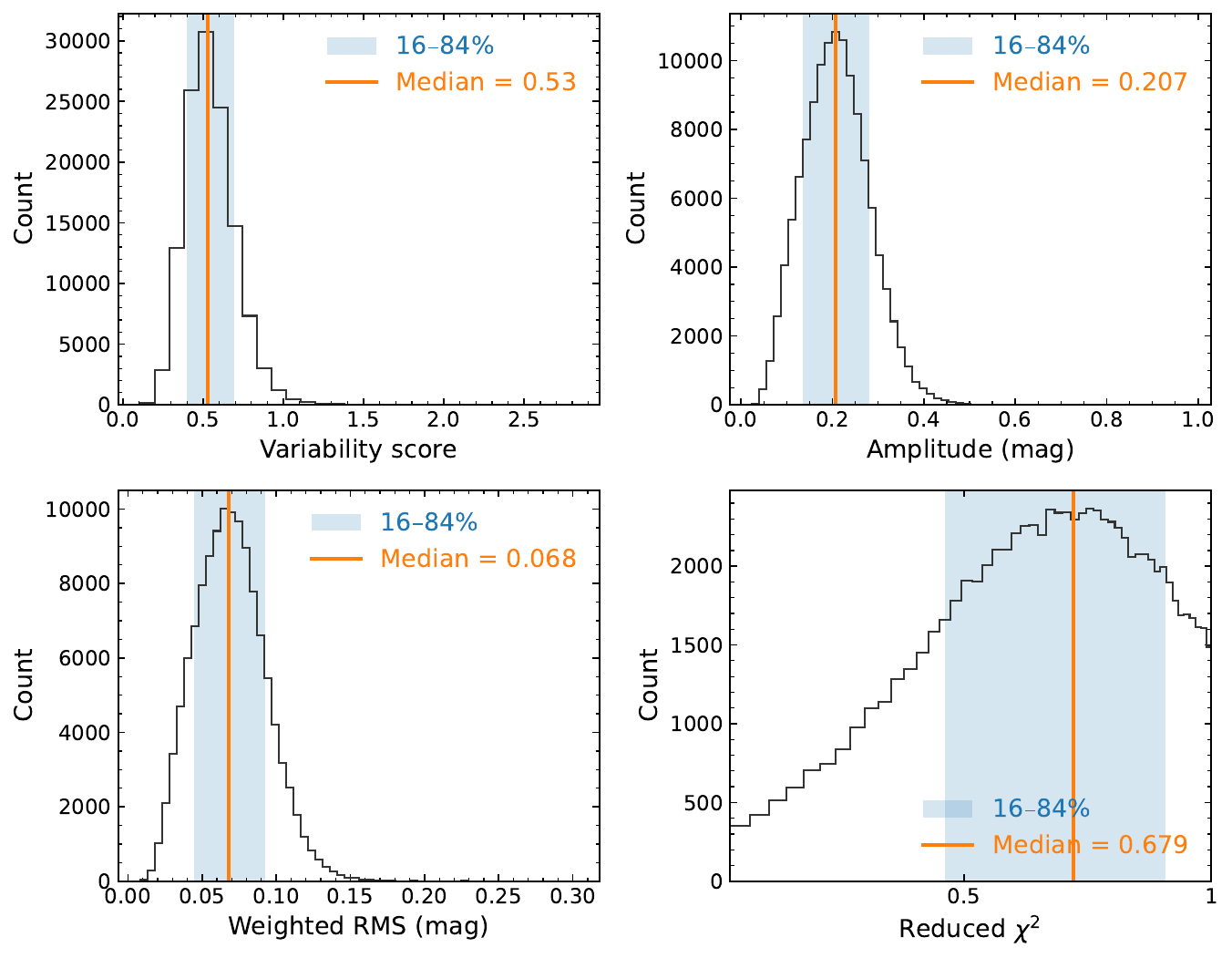}
    \caption{Distributions of mid-infrared variability metrics.
    Clockwise from top left, panels show distributions for the composite variability score, peak-to-peak amplitude, reduced $\chi^{2}$ relative to a constant-flux model, and weighted RMS scatter. In each panel, the vertical orange line marks the median of the distribution, and the shaded region indicates the 16th--84th percentile range. 
    }
    \label{fig:final_variability}
\end{figure*}

To find the most variable sources, we combined the three metrics into one score:
\begin{equation}
S = \mathrm{wRMS} + A + \chi^2_\nu.
\end{equation}
This score goes up with both the amplitude and significance of the variability. We ranked all objects by this score and examined the light curves of the top candidates to verify that the variability was real.  Figure~\ref{fig:final_variability} presents the variability metric distributions for the 99,634 AGN we analyzed, illustrating both the median values of each metric as well as their 16-84\% percentile ranges.

%

\section{Results}
\label{sec:results}

From this process, we investigated the sources with the highest variability scores, emphasizing sources available for spectroscopic follow-up from Palomar Observatory at the time of the analysis. For candidates without archival spectroscopy, Palomar follow-up is essential to confirm the AGN nature of the candidate; a spectroscopic redshift also enables connecting the observed fluxes to inherent luminosities.  For sources with archival spectroscopy, Palomar follow-up enables a search for spectroscopic changes.  
 

This paper reports on two AGN candidates that showed strong mid-infrared flares during the NEOWISE mission, suggestive of mid-infrared selected TDEs:  WISE~J224443.55$+$081606.4 (J2244$+$0816 hereafter) at $z = 0.1185$ and WISE~J215016.59$-$113257.0 (J2150$-$1132 hereafter) at $z = 0.2605$.  Table~\ref{tab:source_properties} lists their basic properties. Based on the variability metric distributions of the 99,634 AGN candidates in the filtered catalog, the two objects of interest are both extreme outliers: J2244$+$0816 lies in the 99.99\% percentile in terms of the combined score, while J2150$-$1132 lies in the 99.98\% percentile. In terms of individual scores, the color variability amplitude of J2244+0816 falls in the 98.51\% percentile, while all other variability metrics for both sources are above the 99.3\% percentile. 

%
\begin{deluxetable}{lcc}
\tablecaption{Properties of the two flare candidates.
\label{tab:source_properties}}
\tablehead{
\colhead{Property} & \colhead{J2244$+$0816} & \colhead{J2150$-$1132}}
\startdata
RA (J2000)       & 22:44:43.55   & 21:50:16.59 \\
Dec (J2000)      & $+$08:16:06.4 & $-$11:32:57.0 \\
Redshift         & 0.1185        & 0.2605 \\
$N_{\rm epochs}$ & 20            & 20 \\  
Score ($S$)      & 21.552        & 19.281 \\      
wRMS (mag)       & 0.125         & 0.213 \\        
Amplitude (mag)  & 0.360         & 0.553 \\   
$\chi^2_\nu$     & 21.067        & 18.514   
\enddata
\end{deluxetable}

\subsection{Lowest Variability Score Targets}
\label{sec:targetsoflow-variability}

Before describing the two identified extremely variable AGN candidates, we take a short digression to discuss R90 candidates with extremely low variability. The R90 AGN catalog is designed with an estimated 10\% contamination rate, assumed due to a combination of inactive galaxies and Galactic (carbon) stars \citep{Assef2018}. Since the former should not be variable, we tested whether the R90 contamination rate depends on the variability score of the targets. 

We selected 9 targets with low variability scores that were observable from Palomar Observatory during a recent observing run in June 2026 and obtained optical spectral data with the New Generation Palomar Spectrograph (NGPS; Fremling et~al., in preparation). Two of the selected targets had archival spectra taken by the Dark Energy Spectroscopic Instrument \citep[DESI;][]{DESI2016}, though we flagged one as having low signal-to-noise data with likely misidentification of features. We obtained spectroscopy for that source and the other seven sources without available spectroscopy. Out of the total sample of 9 selected targets, spectroscopy indicates seven are active galaxies, either with clear broad lines (i.e., type-1 AGN), or narrow emission lines that clearly indicate a type-2 AGN or composite (i.e., active plus inactive) galaxy in the \citet{Baldwin1981} (BPT) line ratio diagram. The remaining two galaxies are normal, star-forming, emission line galaxies according to the BPT diagram. 

We find that 80\% of the low-variability R90 sources are active.  This experiment, albeit based on small-number statistics, suggests that R90 contamination is not concentrated at the lower variability targets as would have been expected had the 10\% contamination been primarily due to misclassified inactive galaxies.  This suggests that the contamination in the R90 catalog is more scattered in terms of their variability properties, as might be expected from Galactic stellar contaminants.  Alternatively, the 10\% contamination rate in the R90 catalog might be an overly conservative estimate.

Table~\ref{tab:low_variability_targets} summarizes the 9 targets observed along with their individual score, wRMS, amplitude, and $\chi^{2}_\nu$ metrics.

\begin{deluxetable}{cccccccc}
\tablecaption{Properties of the low-variability score targets.
\label{tab:low_variability_targets}}
\tablehead{
  \colhead{R.A. (J2000)} & \colhead{Decl. (J2000)} & \colhead{Redshift} &
  \colhead{Classification} & \colhead{$S$} & \colhead{wRMS} &
  \colhead{Amplitude} & \colhead{$\chi^{2}_{\nu}$}
}
\startdata
11:29:49.94 & $+$46:46:28.2 & 1.308 & type-1 AGN             & 0.182 & 0.031 & 0.086 & 0.065 \\
12:08:52.80 & $+$08:42:07.6 & 0.326 & emission-line galaxy  & 0.147 & 0.027 & 0.069 & 0.052 \\
12:24:14.50 & $+$16:31:48.7 & 1.535 & type-1 AGN           & 0.174 & 0.026 & 0.083 & 0.065 \\
12:34:24.89 & $+$65:26:12.8 & 0.520 & type-1 AGN             & 0.220 & 0.009 & 0.024 & 0.187 \\
13:22:32.23 & $+$42:57:26.3 & 1.487 & type-2 AGN             & 0.173 & 0.028 & 0.083 & 0.062 \\
14:13:49.46 & $+$39:44:48.5 & 1.188 & type-1 AGN             & 0.205 & 0.033 & 0.088 & 0.084 \\
17:36:43.34 & $+$60:52:21.4 & 0.329 & composite galaxy       & 0.221 & 0.017 & 0.044 & 0.160 \\
17:37:46.78 & $+$46:56:56.4 & 0.172 & type-2 AGN             & 0.213 & 0.031 & 0.099 & 0.083 \\
22:10:09.98 & $+$28:15:18.7 & 0.187 & emission-line galaxy   & 0.212 & 0.033 & 0.086 & 0.092
\enddata
\tablecomments{The first eight sources were observed on the nights of UT 2026 June 13 and 15 with NGPS; these were both photometric nights with seeing of $\simeq 1.4^{\prime \prime}$.  Exposure times ranged between 300~s and 1200~s.  See Section~4.3 for details on the instrument configuration and data analysis, though note that all four channels of NGPS were available in June 2026. The redshift and classification of the final source is based on DESI data.  J141349.46+394448.5 was misidentified as a galaxy at $z = 0.5377 \pm 0.0001$ by DESI with {\tt ZWARN} = 0, though, in retrospect, the DESI spectrum shows the broad \ion{C}{3}] and self-absorbed broad \ion{C}{4} of a (type-1) quasar at $z = 1.188$.}
\end{deluxetable}


\subsection{Multi-wavelength Light Curves}
\label{sec:lightcurves}

\begin{figure*}[t]
    \centering
    \includegraphics[width=1\textwidth]{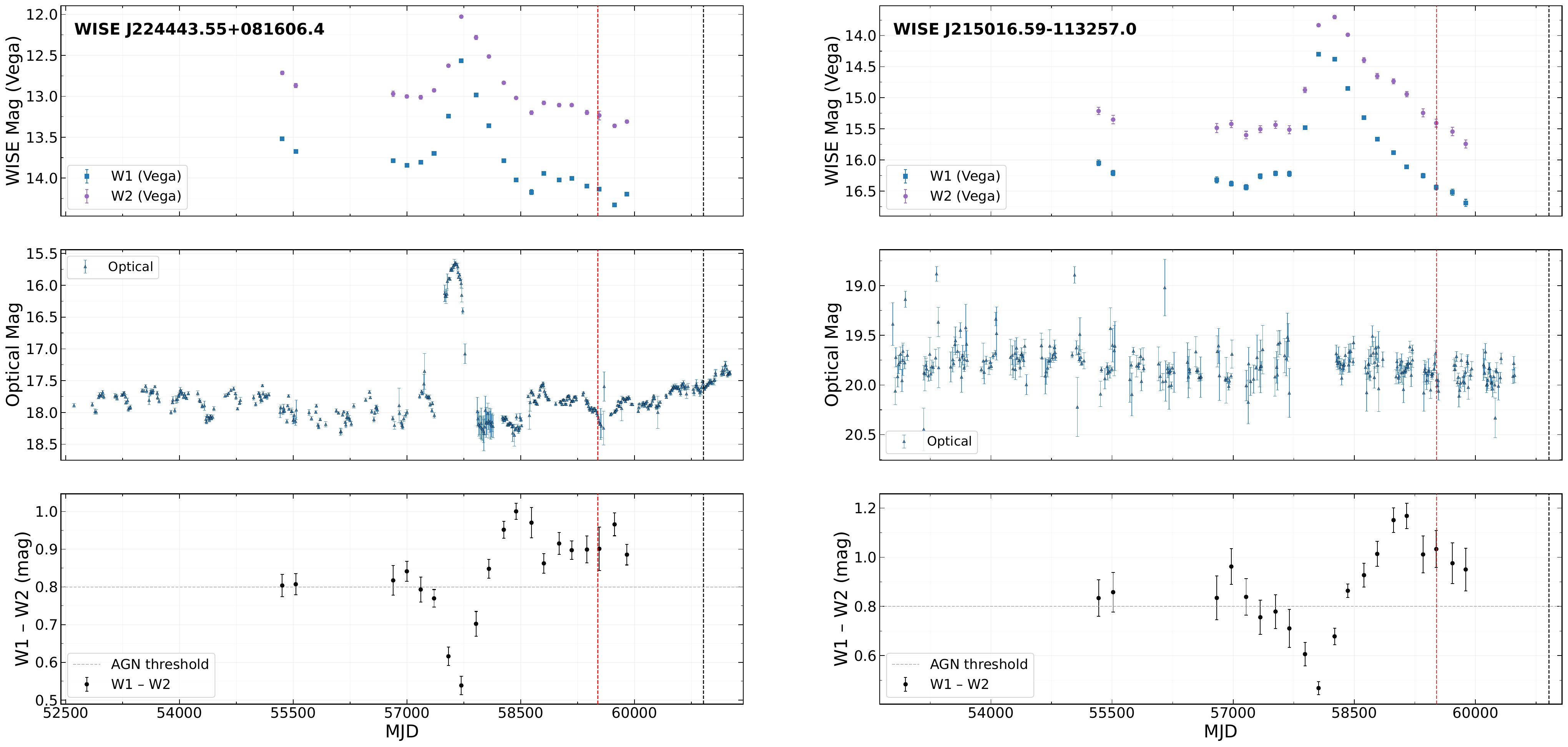}
    \caption{\emph{\textbf{Left}}: Light curve of WISE J2244$+$0816 showing the mid-infrared and optical variability behavior. \textit{Top:} WISE W1 (3.4\,\micron) and W2 (4.6\,\micron) epochal photometry from the AllWISE and NEOWISE missions, sampled at the standard $\sim$6-month cadence. \textit{Middle:} Optical data from eight surveys, co-registered to a common $r$-band scale and plotted with a single symbol, illustrating a significant flare almost simultaneous with the mid-infrared flare. \textit{Bottom:} Epochal W1$-$W2 color evolution, highlighting the mid-infrared response of color during the flare and the subsequent settling of the color to a redder state compared to the pre-flare baseline. The dashed horizontal line in the bottom panel illustrates the \citet{Stern2012} threshold for separating likely AGN from quiescent galaxies.  Vertical lines indicate dates of optical spectra.
    \emph{\textbf{Right}}: Similar to the plot on the left but for J2150$-$1132. J2150$-$1132 doesn't exhibit flaring behaviour in the optical wavebands.}
    \label{fig:Plot_J2244_0816}
\end{figure*}

Returning to the highly variable sources, we present the light curves of J2244$+$0816 and J2150$-$1132 in Figure~\ref{fig:Plot_J2244_0816}. We find that the source J2244$+$0816 (left panel) exhibits typical low-level mid-infrared AGN variability from 2010 to early 2016, at which point it brightened by $\sim 1.5$~magnitudes in both WISE bands. The flare peaked around MJD 57700 (late 2016) and has been fading since, with the latest flux measurements fainter than the pre-flare flux. The optical photometry combines data from Pan-STARRS1 $r$-band, ZTF Forced Photometry (ZFOR) and ZTF $r$-band, the Catalina Sky Survey (CSSF) and Mount Lemmon Survey (MLSF) $r$-band, LINEAR $r$-band, ATLAS $o$-band, and ASAS-SN $V$-band, all scaled to a common $r$-band scale. The combined light curve spans MJD 52609 to 61261, providing roughly seven years of coverage before the mid-infrared flare. We show that the optical data exhibits significant flaring activity. Given the 6-month cadence of WISE, Figure~\ref{fig:Plot_J2244_0816} implies a time lag of $\leq$~105 days between the optical and mid-infrared flares. The bottom panel shows the W1$-$W2 color variability. Before the flare, the mid-infrared color was steady at about 0.8~mag, just above the \citet{Stern2012} threshold for separating active and quiescent galaxies. During the flare rise, the source became bluer, dropping to $\sim 0.55$~mag, nominally indicative of an inactive galaxy. After the flare peak, the source became redder during the decline, peaking at W1$-$W2 $\sim 1.0$, before ultimately stabilizing at W1$-$W2 $\sim 0.9$, which is slightly redder than the pre-flare mid-infrared color.

The light curve of J2150$-$1132 is shown in Figure~\ref{fig:Plot_J2244_0816}. The flare from this source was even more intense, brightening by about 2 magnitudes in both W1 and W2 between 2016 and 2018. The flare peaked around MJD 58200 (early 2018) and has since faded, again with the latest post-flare data fainter than the pre-flare brightness. Unlike J2244$+$0816, the combined optical light curve shows little to no variability. Overall, the optical and mid-infrared data suggest the optical light is dominated by the host galaxy in the observed data, with an active galactic nucleus hidden by dust along our line-of-sight (i.e., a type-2 AGN). The W1$-$W2 color evolution of J2150$-$1132 follows a similar pattern to J2244$+$0816: the source initially showed mid-infrared AGN colors, then became bluer during the rise, dropping to about 0.45 mag, and then redder during the decline, reaching above 1.1 mag before ultimately stabilizing with mid-infrared photometry slightly fainter and redder than pre-flare.


This color behavior matches the predictions of the mid-infrared-emitting dusty torus responding to activity from the accretion disk, i.e., a dust echo as described by \citet{vanVelzen2016}, and \citet{masterson2024newpopulationmidinfraredselectedtidal}. The bluer mid-infrared color observed during the early flare is predicted to be the result of the hotter dust in the inner portions of the dusty torus and closer to the black hole exhibiting an initial reaction to the nuclear transient. The later reddening occurs both as the light echo reaches farther into the dusty region, illuminating cooler material and as the inner, initially heated material fades. The lack of optical variability in J2150$-$1132 suggests this event was heavily obscured, with the mid-infrared coming from dust reprocessing a hidden UV/optical flare.

\subsection{Optical Spectroscopy}
\label{sec:spectroscopy}

We obtained new optical spectroscopic observations of both sources with NGPS on the 200-inch Hale Telescope at Palomar Observatory to look for spectral changes compared to archival spectral data available from the DESI survey.  

We observed J2244$+$0816 for 300~s on UT 2025 August 23 and we observed J2150$-$1132 for 900~s on UT 2025 August 24. Both nights suffered from intermittent light cirrus and the seeing ranged over 1.2-1.5$^{\prime\prime}$.  We used the 1.5$^{\prime\prime}$ wide slit for these longslit observations and at the time of these observations, only the $r$- and $i$-channels of NGPS were available.  We used the $2 \times 2$ binning mode for the readout of both channels, which provides $R \equiv \lambda / \Delta \lambda \sim 1500$ spectral resolution from 5900~\AA\ to $\sim 1.04\, \mu$m.  We reduced the NGPS spectra using standard techniques within IRAF and flux calibrated the spectra using observations of standard stars from \citet{Massey1990} observed during the same observing run.

\subsubsection{J2244$+$0816}

\begin{figure*}[t]
    \centering
    \label{opticalspectraJ2244}
    \includegraphics[width=0.9\textwidth]{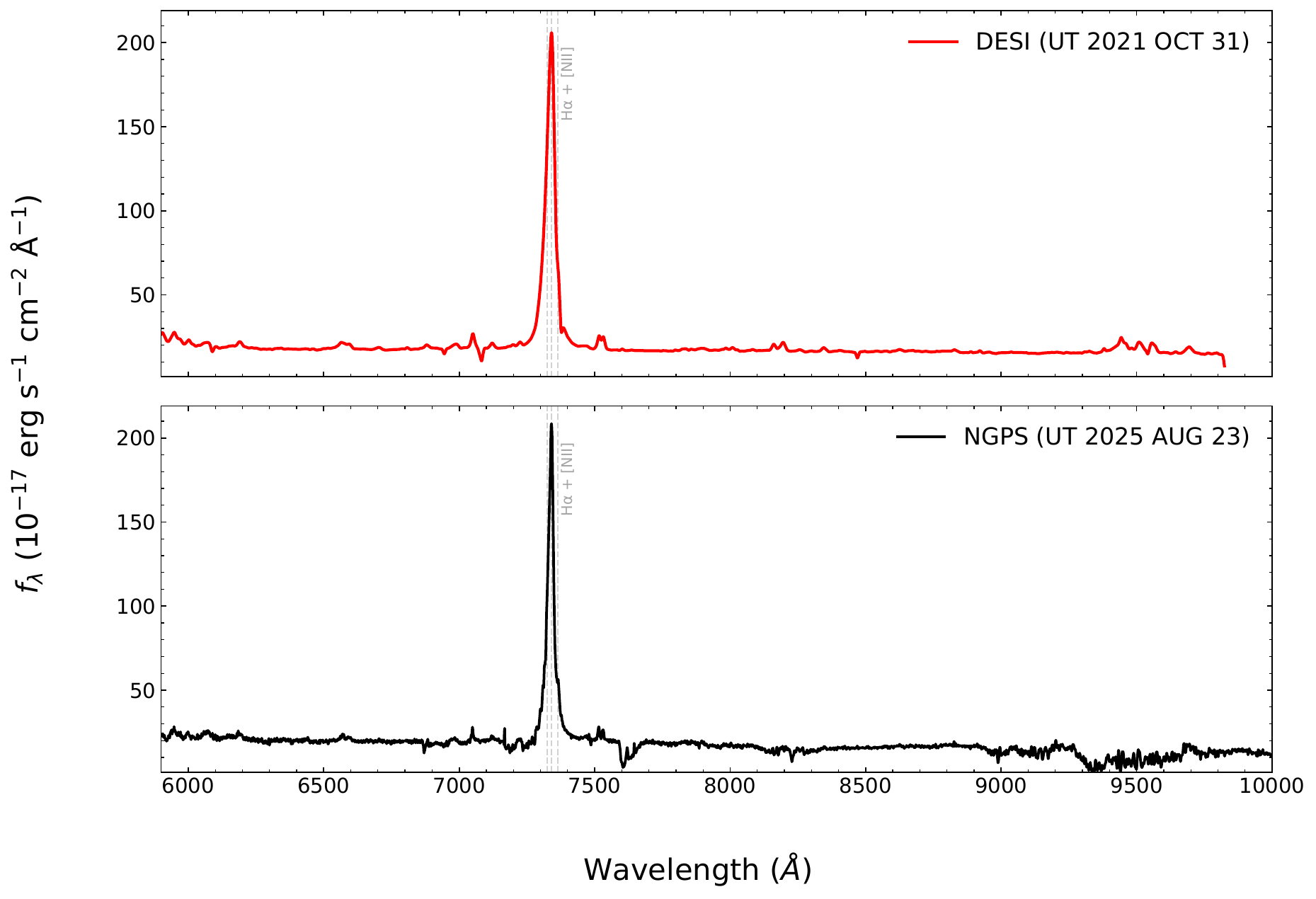}
  
    \caption{Optical spectroscopy of J2244$+$0816. \textbf{Top:} Archival DESI spectrum obtained on UT 2021 October 31. \textbf{Bottom:} Follow-up Palomar NGPS spectrum from UT 2025 August 23. Both spectra are dominated by the strong, somewhat broadened H$\alpha$, with weaker narrow [N\,{\sc ii}] and [S\,{\sc ii}] doublet lines evident. There is no strong evidence of spectroscopic changes between the two observations.
    }
    \label{fig:spec_J2244}
\end{figure*}

We compare the spectra of J2244$+$0816 from two different epochs in Figure~\ref{fig:spec_J2244}. The DESI spectrum was taken on UT 2021 October 31, about five years after the mid-infrared flare peak, while the Palomar spectrum was taken approximately four years later.

Both spectra reveal clear, somewhat broadened H$\alpha$ at redshift $z = 0.1185$, as well as the [S\,{\sc ii}]\,$\lambda\lambda$6716,6731 doublet, and indications of the [N\,{\sc ii}]\,$\lambda\lambda$6548,6584 doublet. The NGPS H$\alpha$ emission line, which is the strongest feature, is modestly broadened with a velocity width of $1240\, {\rm km}\, {\rm s}^{-1}$ when fit with a single line. Although this is a slight overestimate since the weak flanking [N\,{\sc ii}] lines were not included, the H$\alpha$ width is still indicative of accretion onto an SMBH. Without pre-flare spectra available, it is unclear if the observed spectrum is related to a newly formed accretion disk; e.g., accretion from TDEs often continues at a low rate for several years \citep[e.g.,][]{Komossa2015}. 

The H$\beta$ + [O\,{\sc iii}] region was not covered by the Palomar data.  However, the DESI data does cover that range and shows a broadened H$\beta$ line with a measured velocity width of $1500\, {\rm km}\, {\rm s}^{-1}$, similar to the H$\alpha$ line width, and indicating an active galaxy.  The H$\beta$ and [O\,{\sc iii}]$\lambda 5007$ lines have comparable strengths.

\subsubsection{J2150$-$1132: A Changing-Look AGN}

\begin{figure*}[t]
    \centering
    \includegraphics[width=0.9\textwidth]{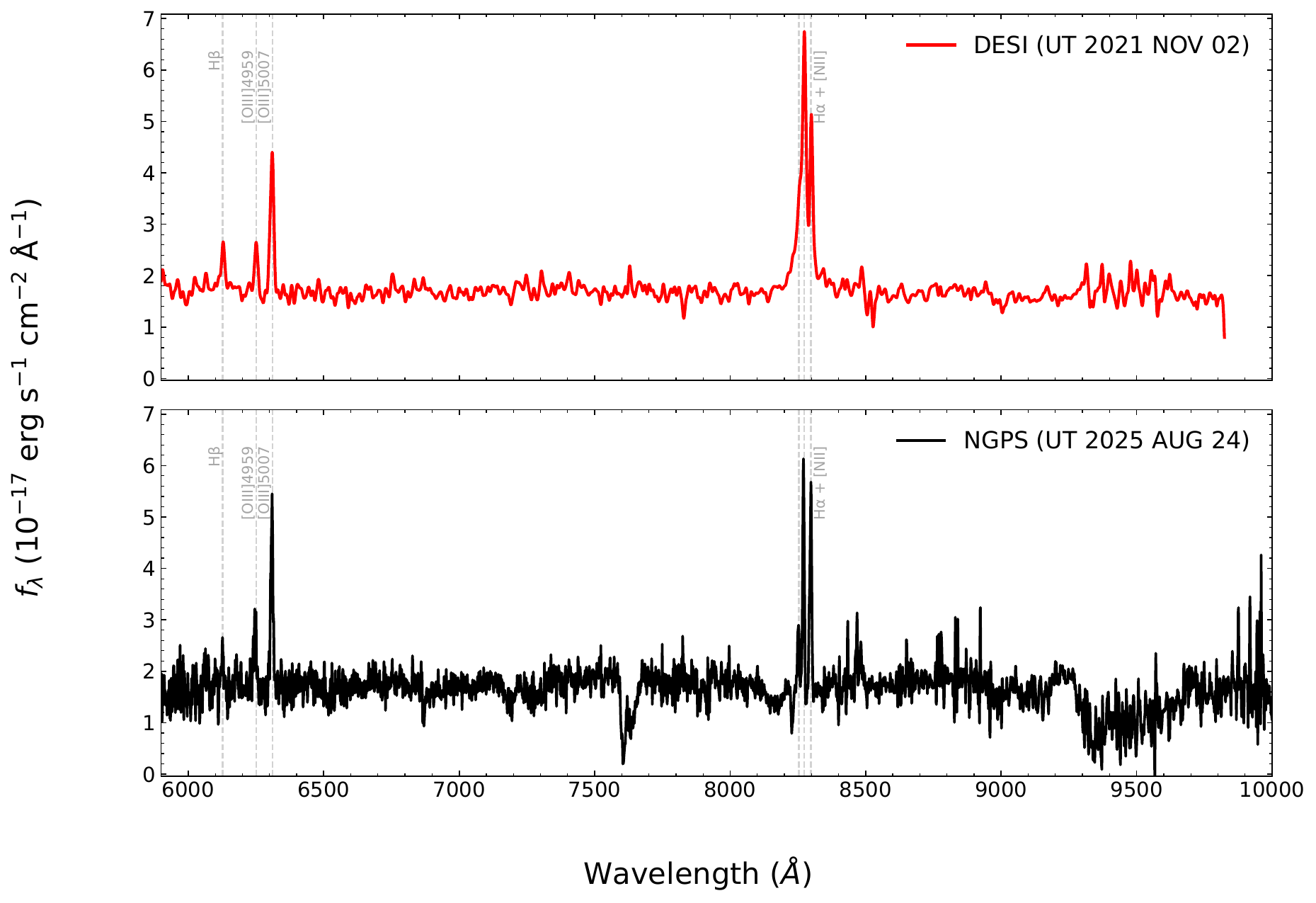}

    \caption{Optical spectroscopy of J2150$-$1132. \textbf{Top:} Archival DESI spectrum obtained on UT 2021 November 2. \textbf{Bottom:} Follow-up Palomar NGPS spectrum from UT 2025 August 24. Both spectra show strong, narrow emission from [O\,{\sc iii}] and [N\,{\sc ii}], as well as H$\alpha$ emission which weakens and becomes narrower over time.  H$\beta$ is weakly evident in the DESI spectrum.}
    \label{fig:spec_J2150}
\end{figure*}

We compare the two spectra of J2150$-$1132 in Figure~\ref{fig:spec_J2150}. The DESI spectrum was taken on UT 2021 November 2, about three years after the mid-infrared flare peak, while the Palomar spectrum was taken approximately four years later. In contrast to J2244$+$0816, we find clear differences between the two J2150$-$1132 spectra.

The DESI spectrum clearly shows several emission lines at $z = 0.2605$, namely H$\beta$, [O\,{\sc iii}], H$\alpha$, and [N\,{\sc ii}].  The H$\alpha$ line is broadened, while the H$\beta$ line is narrow, indicating a type~1.9 AGN, i.e., an obscured AGN.  The [N\,{\sc ii}]/H$\alpha$ vs.\ [O\,{\sc iii}]/H$\beta$ line ratios also indicate an active galaxy based on the BPT diagram.

Four years later, Palomar data reveals clear spectral changes, with all the emission lines now narrow.  The H$\alpha$ in these data has a width of $200\, {\rm km}\, {\rm s}^{-1}$, comparable to the other emission lines, and H$\beta$ is no longer evident. The line ratios clearly indicate a Seyfert AGN \citep[e.g.,][]{Baldwin1981}. This kind of change, where broad lines fade and only narrow lines remain, indicates a changing-look or changing-state AGN \citep[e.g.,][]{LaMassa2015, MacLeod2016, Graham2020, Ricci2020}. 

\subsection{Multiwavelength Properties}
\label{sec:multiwavelength}

J2244$+$0816 is detected at 1.4~GHz by the Faint Images of the Radio Sky at Twenty-Centimeters (FIRST) survey \citep{Becker1995} as a marginally resolved 5.01~mJy source with a deconvolved extent of 1.28$^{\prime\prime}$.  This flux density corresponds to a radio luminosity of $1.9 \times 10^{23}\, {\rm W}\, {\rm Hz}^{-1}$, which is below the traditional threshold used to identify radio-loud AGN.  The source was also detected by ROSAT in the latter half of 1990 as 2RXS~J224443.6+081605 with $32.9 \pm 6.3$ counts \citep{Boller2016}. This corresponds to an X-ray flux of $2.9 \times 10^{-12}\, {\rm erg}\, {\rm cm}^{-2}\, {\rm s}^{-1}$, or a typical low-luminosity AGN X-ray luminosity of $10^{44}\, {\rm erg}\, {\rm s}^{-1}$.

J2150$-$1132 has no reported archival radio or X-ray detections.

\subsubsection{XMM-Newton Observations}

We observed WISE J2244+0816 on 2026 June 4 with XMM-Newton to study the current accretion state of the system (obsID 0982950101). We used {\sc xmmsas} v22.1.0 to analyze the data \citep{gabeil+04}. We first checked for periods of high background by creating a lightcurve of the events from the entire detector in the 10--12 keV band, finding a strong flare towards the towards the end of the observation seen in the pn, MOS1 and MOS2 detectors. We filter the events to exclude periods where the 10--12 keV rate is greater than 0.4 counts\,s$^{-1}$ in the pn, and greater than 0.35 counts\,s$^{-1}$ in the MOS detectors. This leaves 4.8 ks of exposure in the pn, and 6.9 ks of exposure in the MOS detectors.  Events were selected with {\tt PATTERN $\leq4$} for the pn and {\tt PATTERN $\leq12$} for the MOS. A circular region with a radius of 20\arcsec\ was used to extract the source spectrum and an annular region with an inner radius of 50\arcsec\ and outer radius of 80\arcsec\ was used to extract the background spectrum. Data from the pn and both MOS instruments were extracted in this way. 

The source was well detected with a count rate of $0.75\pm0.01$ counts\,s$^{-1}$ in pn, $0.18\pm0.01$ counts\,s$^{-1}$ in MOS1, and $0.16\pm0.01$ counts\,s$^{-1}$ in MOS2 in the 0.2--10 keV energy range. All spectra were grouped with a minimum of one count per bin using the {\sc heasoft} tool {\tt grppha} and fitted in {\sc xspec} \citep{KArnaud}. We used the Cash, or C-statistic \citep{cash79} to fit the background-subtracted spectra. Since the C-statistic formally cannot be used when the background is subtracted, {\sc xspec} uses a modified version of the C-statisitic known as the W-statistic to account for this.

We fit the pn, MOS1 and MOS2 spectra simultaneously with a cross-correlation constant to account for calibration uncertainties, fixing the pn constant to unity. We first tried an absorbed power-law subjected to absorption in our Galaxy, {\tt TBabs}, where $N_{\rm H}=6\times10^{20}\, {\rm cm}^{-2}$ from the HI4PI survey \citep{HI4PI}, with additional absorption intrinsic to the source at $z=0.118$, {\tt zTBabs}. The resultant power-law index is steep with $\Gamma=3.42 \pm 0.11$ and we find no evidence for absorption in addition to the Galactic value. From a statistical perspective, the model fits reasonably well with $C=1489.13$ for 1286 degrees of freedom, though strong spectral residuals suggest the presence of a soft excess component. We next tried adding a {\tt cutoffpl} component, which improved the fit to $C=1195.18$ with 1283 degrees of freedom. Here $\Gamma_{\rm pl}=2.34^{+0.14}_{-0.15}$, $\Gamma_{\rm cutoffpl}=-4.14^{+0.17}_{-0.12}$ and $E_{\rm cutoff}=8_{-2}^{+0.01}\times10^{-2}$ keV. Since the latter are unphysical values we also tested a {\tt diskbb} model for the soft excess. This yielded $C=1202.20$ with 1284 degrees of freedom where $\Gamma_{\rm pl}=2.43\pm0.15$, $T_{\rm in}=0.11\pm0.01$ keV. We adopt this model as the best fit to the data. The observed 0.2--10 keV flux is $1.3_{-0.8}^{+0.5}\times10^{-12}\, {\rm erg}\, {\rm cm}^{-2}\, {\rm s}^{-1}$ which implies a luminosity of $4.6_{-2.8}^{+1.8}\times10^{43}\, {\rm erg}\, {\rm s}^{-1}$, or roughly a factor of three drop in flux compared to the pre-flare ROSAT observation.

%
%

\section{Dust Echo Modeling}
\label{sec:modeling}

Having established that both sources flared strongly in the mid-infrared, we now model those flares to measure three physical quantities: the temperature of the emitting dust, the radius at which it sits, and the total energy the flare radiated.  Our approach throughout assumes that the mid-infrared emission is a dust echo, in which UV and optical light from an event near the black hole is absorbed by surrounding dust and re-emitted at longer wavelengths \citep{vanVelzen2016, masterson2024newpopulationmidinfraredselectedtidal}.Under that assumption, the mid-infrared light curve is a smoothed copy of the intrinsic flare, broadened by the time it takes light to cross the dusty region.

We proceed in three steps.  First we isolate the flare component of the observed flux by subtracting the underlying AGN and host emission (Section~\ref{sec:baseline}).  We then use the W1/W2 color of the flare to measure how the dust temperature evolves (Section~\ref{sec:temperature}). Finally, we fit the flare light curves with a spherical dust shell model to measure the dust radius and the disruption time (Section~\ref{sec:dustshell}). The temperature measurement and the shell fit are independent of one another, except that the shell fit adopts the flare-averaged temperature to relate the W1
and W2 amplitudes.

\subsection{Isolating the Flare Flux}\label{sec:baseline}

Both sources are active galaxies, so the observed mid-infrared flux at any epoch is the sum of a steady component from the AGN and host galaxy and a transient component from the flare. We refer to the latter as the \emph{flare flux}, $F_{\rm flare} = F_{\rm obs} - F_{\rm base}$, and it is the quantity modeled in the rest of this section.

Choosing $F_{\rm base}$ is less straightforward here than in previous work modeling TDEs in quiescent galaxies. Such studies generally average all pre-flare epochs \citep[e.g.,][]{vanVelzen2016, vanVelzen2021a, Jiang2021}; for example, \citet{masterson2024newpopulationmidinfraredselectedtidal} subtracts the pre-flare reference co-add at the image level.  Such an approach, however, is not appropriate for an active galaxy for which the underlying AGN varies on its own. We therefore adopt four baselines and compare them: (1) the mean of the final three epochs, when the flare has faded and the source is closest to quiescence; (2) the single faintest epoch; (3) the mean of the pre-flare epochs; and (4) no subtraction at all. We adopt the post-flare mean as our primary baseline, since it best represents the quiescent AGN state.  For J2244$+$0816 this gives $F_{\rm base} = 0.637 \pm 0.072$\,mJy in W1 and $0.822 \pm 0.087$\,mJy in W2; for J2150$-$1132, his gives $F_{\rm base} =0.075 \pm0.009$\,mJy and $0.103 \pm 0.014$\,mJy.


The remaining baselines define the systematic uncertainty this choice introduces, and that uncertainty is modest.  Considering the total radiated energy of the flare, the key measurement from this modeling, the post-flare mean gives the lowest, most conservative flare energy, while other baselines increase that flare energy by $\leq 0.23$~dex for J2244$+$0816 and by $\leq 0.43$~dex for J2150$-$1132.

We convert WISE magnitudes to flux densities using zero points of 309.540\,Jy for W1 and 171.787\,Jy for W2 \citep{Wright2010}, and add a 10\% systematic uncertainty in quadrature to account for calibration errors, as recommended by \citet{Christos2023} and  \citet{masterson2024newpopulationmidinfraredselectedtidal}. All luminosities and energies below are computed in the rest frame, adopting $L_\nu = 4\pi D_L^2 f_\nu / (1+z)$, where $D_L$ is the luminosity distance for our adopted cosmology.

\subsection{Dust Temperature Evolution}\label{sec:temperature}

With photometry in only two bands, the W1/W2 flux ratio at each epoch gives a single color temperature. We fit each epoch with both a blackbody and a modified blackbody (graybody) with emissivity $\propto \nu^{1.8}$, as appropriate for graphite dust \citep{Draine1984}.  Uncertainties come from Monte Carlo resampling: at each epoch we draw 1000 realizations of the two fluxes from their errors, refit the temperature each time, and take the 16th, 50th and 84th
percentiles.

Not every epoch bounds the temperature equally well, so we classify each by the constraint it provides.  Epochs detected in both bands give a two-sided constraint; epochs detected in only one give a one-sided constraint, which still carries real information and which we retain; and epochs whose fitted temperature runs into the upper edge of the fitting grid at 2500~K, or which the Monte Carlo simulations do not constrain consistently, are excluded from the flare-averaged
values.

\begin{figure*}[t]
    \centering
    \includegraphics[width=0.85\textwidth]{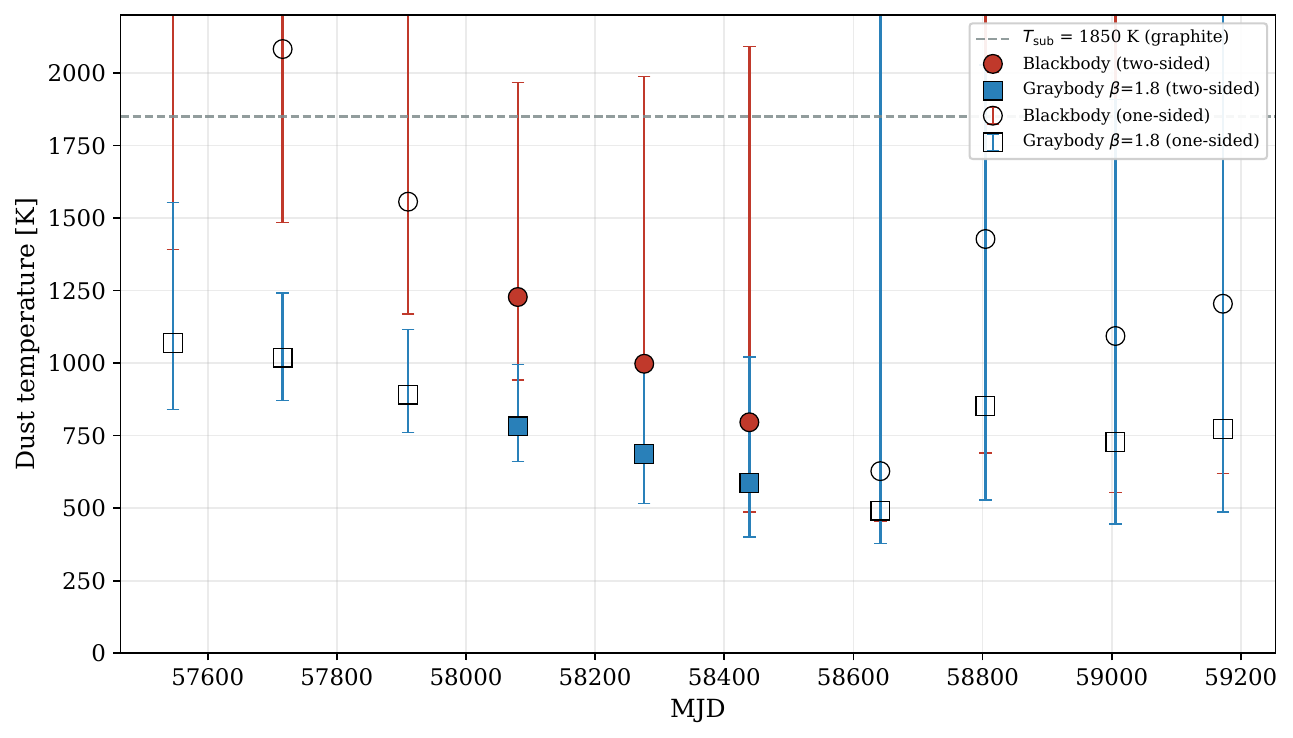}
    \caption{Dust temperature evolution of J2244$+$0816 derived from the W1/W2 flux ratios of the baseline-subtracted photometry.  Circles show blackbody fits and squares show graybody ($\beta = 1.8$) fits; filled symbols mark the three epochs with two-sided temperature constraints, open symbols show the seven epochs for which only one band is detected and the constraint is one-sided.  Error bars give the 16th--84th percentile range from 1000 Monte Carlo resamplings of the photometric errors.  The horizontal dashed line marks the graphite sublimation temperature, $T_{\rm sub} = 1850$~K.  The flare-averaged blackbody temperature is $\langle T_{\rm BB}\rangle = 1336 \pm 171$~K.  The one-sided epochs at MJD 57546--57910 carry blackbody temperatures above $T_{\rm sub}$ with lower bounds extending below it, and are unconstrained at the upper end; they are retained in the average as lower limits but do not by themselves indicate dust hotter than sublimation.}
    \label{fig:diag_J2244}
\end{figure*}

Figure~\ref{fig:diag_J2244} shows the result for J2244$+$0816.  Ten epochs (7--16) fall within the flare window, MJD 57546--59172, of which three are two-sided and seven one-sided.  The blackbody temperature falls from 2350~K at the first epoch of the window to 1205~K at the last, though the early values are poorly constrained one-sided fits; averaged over the window, $\langle T_{\rm BB}\rangle = 1336 \pm 171$~K, with a systematically cooler $\langle T_{\rm GB}\rangle = 788$~K. 

\begin{figure*}[t]
    \centering
    \includegraphics[width=0.85\textwidth]{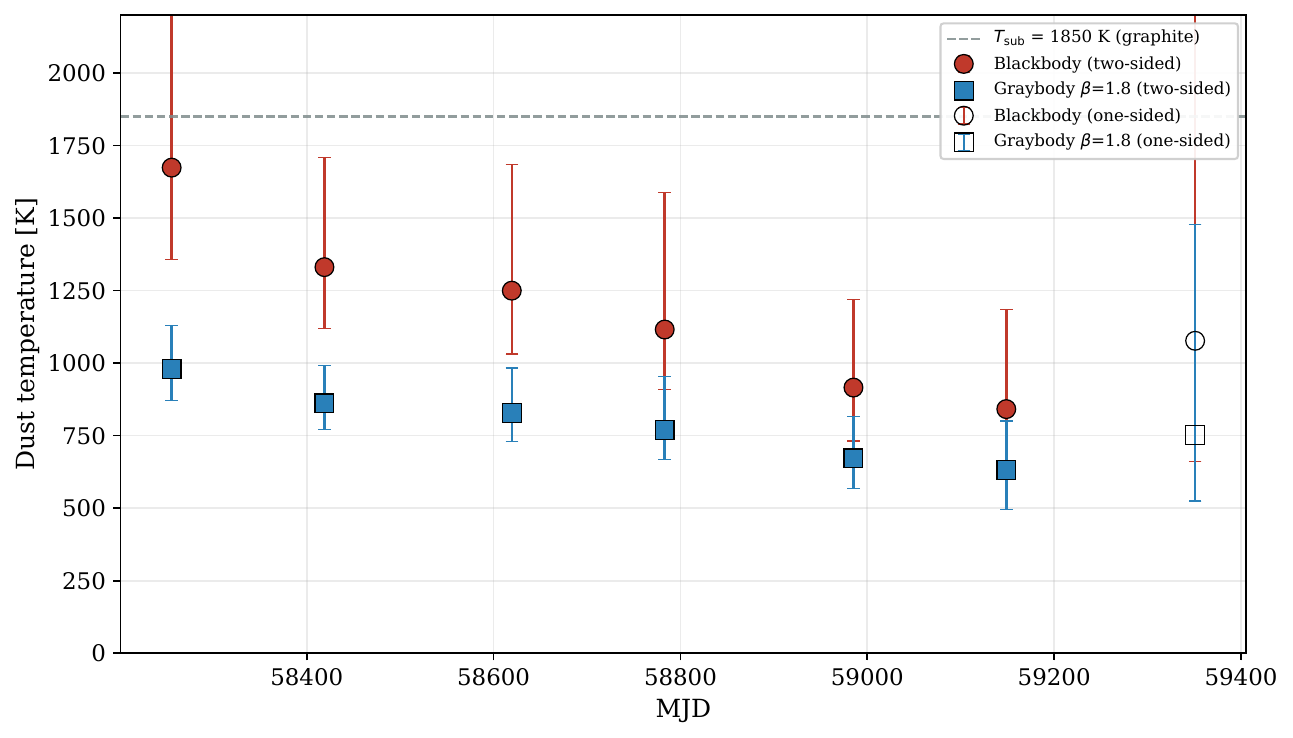}
    \caption{Same as Figure~\ref{fig:diag_J2244}, but for J2150$-$1132, for which six of the seven epochs in the flare window give two-sided constraints.  The blackbody temperature declines monotonically from $1674$~K at MJD 58255 to $841$~K at MJD 59149, and the graybody temperature from $980$~K to $631$~K, over roughly 900 days.  This smooth cooling is the signature expected of a dust echo as the reprocessing region is illuminated at progressively larger radii and the inner, initially heated material fades.  The flare-averaged blackbody temperature is $\langle T_{\rm BB}\rangle = 1172 \pm 106$~K.  Note that epochs 9 and 10, which bracket the flare peak, are excluded because their fitted temperature reaches the 2500~K ceiling of the fitting grid.}
    \label{fig:diag_J2150}
\end{figure*}

Figure~\ref{fig:diag_J2150} shows J2150$-$1132, for which the window MJD 57890--59351 contains seven usable epochs (11--17), six of them two-sided.  Here the temperature is much better determined: it declines monotonically from 1674K at MJD 58255 to 841K at MJD 59149, giving $\langle T_{\rm BB}\rangle = 1172 \pm106$~K and $\langle T_{\rm GB}\rangle = 785$~K. 

Two epochs of J2150$-$1132, 9 and 10, are excluded because their fitted temperature reaches the 2500~K grid ceiling.  These are the brightest epochs of the flare (S/N of 8.6 and 7.5 in W1), so their exclusion is not a signal-to-noise issue: their W1/W2 color simply implies a temperature above the graphite sublimation limit, and therefore cannot be reprocessed dust emission.  Because both lie on the rising phase, the energy integral in Section~\ref{sec:energetics} covers only the decay. We return to this point
below.

In both sources, the retained temperatures lie below the graphite sublimation temperature, $T_{\rm sub} \approx 1850$~K \citep{vanVelzen2016, 2012MNRAS.420..526M}, with the hottest epochs approaching it.  This is what a dust echo should look like: the flare destroys any dust inside the sublimation radius and heats the dust just outside it to near-sublimation temperatures, which then cools as the flare fades.  We note that these temperatures follow from assuming a single dust temperature at each epoch. In reality, the emission sums over a range of dust temperatures and radii, so the values are characteristic rather than exact.


\subsection{Spherical Dust Shell Model}\label{sec:dustshell}

To measure the dust radius, we fit the flare light curves with the spherical shell model of \citet{vanVelzen2016, vanVelzen2021a}, which was also applied by \citet{masterson2024newpopulationmidinfraredselectedtidal}.  The model places the dust in a thin spherical shell of radius $R_{\rm dust}$ around the black hole and assumes the intrinsic flare follows the
canonical fallback rate,
\begin{equation}
L_{\rm int}(t) = L_0 \left( \frac{t - t_{\rm p} + t_0}{t_0} \right)^{-5/3},
\end{equation}
where $t_{\rm p}$ is the time of peak luminosity, $t_0$ is the fallback timescale, and $L_0$ is the peak intrinsic luminosity.  Because we cannot measure the intrinsic UV/optical flare in either source as J2244$+$0816 has only sparse optical coverage, and J2150$-$1132 has no detected optical counterpart at all, $L_0$ is not constrained by our data.  We therefore fix $L_0 = 1$ and fit
only the shape of the light curve, absorbing the overall normalization into a separate amplitude parameter for each band, following \citet{masterson2024newpopulationmidinfraredselectedtidal}.

The observed mid-infrared light curve is this intrinsic flare convolved with the response of the shell,
\begin{equation}
L_{\rm IR}(t) = \int_0^{2R_{\rm dust}/c} \Psi(\tau)\, L_{\rm int}(t - \tau)\, d\tau,
\end{equation}
where $\tau$ is the light travel delay between the flare and the observer, running from zero for dust on the near side of the shell to $2R_{\rm dust}/c$ for dust on the far side, and $\Psi(\tau)$ is a top-hat function of that width. The effect is to smear the intrinsic flare over the light-crossing time of the shell, so a broader observed light curve implies a larger dust radius.

We fit both bands simultaneously using the MCMC sampler \texttt{emcee} \citep{ForemanMackey2013}, with 32 walkers, 500 burn-in steps and 5000 production steps.  The free parameters are $R_{\rm dust}$, the W1 amplitude, $t_{\rm p}$ and $t_0$; the W1/W2 amplitude ratio is fixed by the flare-averaged blackbody temperature from Section~\ref{sec:temperature}, giving $L_{\rm W2}/L_{\rm W1} =
1.066$ for J2244$+$0816 and $1.357$ for J2150$-$1132.

\begin{figure*}[t]
    \centering
    \includegraphics[width=0.85\textwidth]{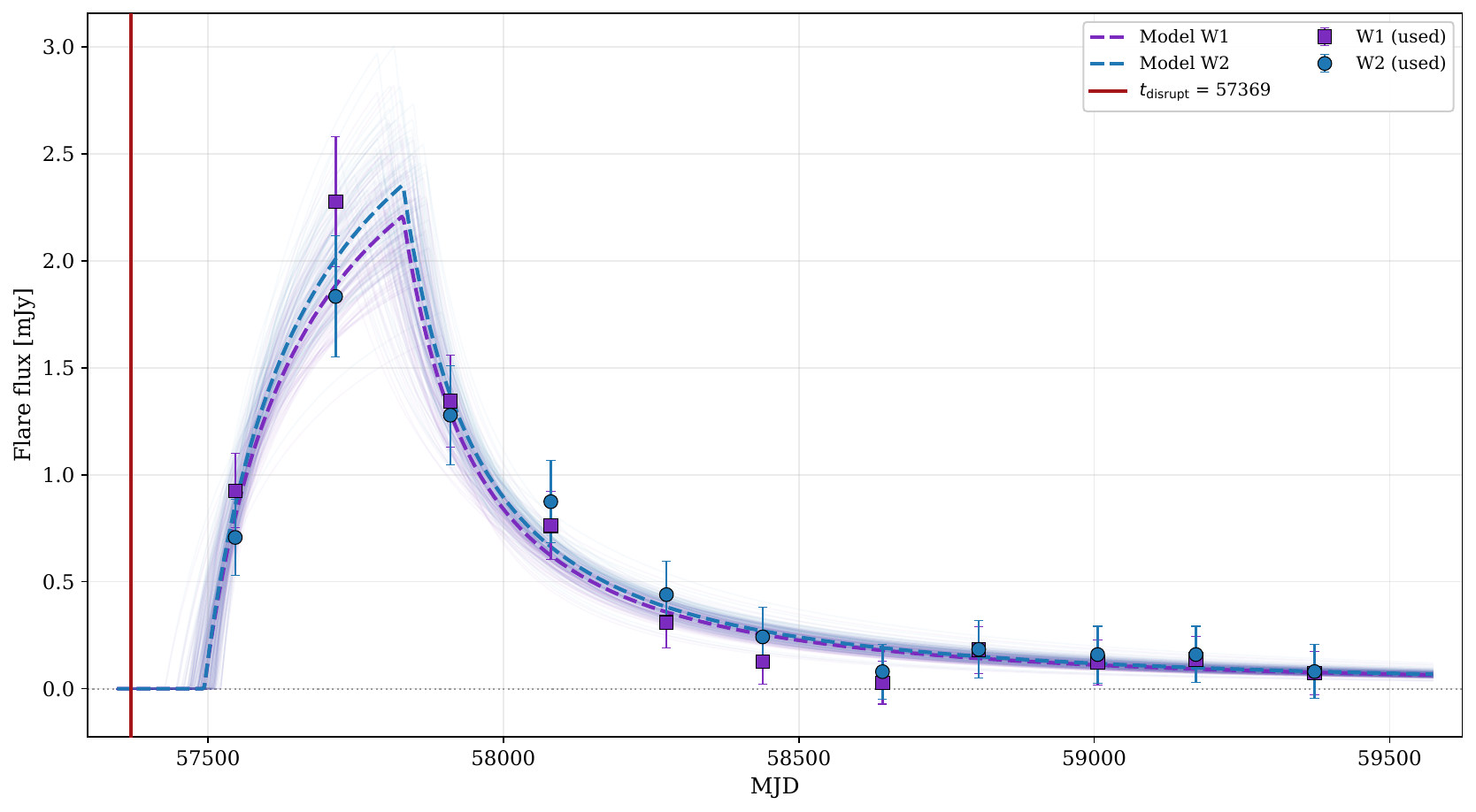}
    \caption{WISE W1 and W2 flux difference light curves for J2244+0816 fitted with a simple spherical dust model. Dashed lines give the best fit lines in both W1 and W2 bands. The shaded region shows 100 posterior draws from the MCMC chain. The solid vertical line indicates $t_{\rm disrupt}$.}
    \label{fig:echomodel_J2244}
\end{figure*}

\begin{figure*}[t]
    \centering
    \includegraphics[width=0.85\textwidth]{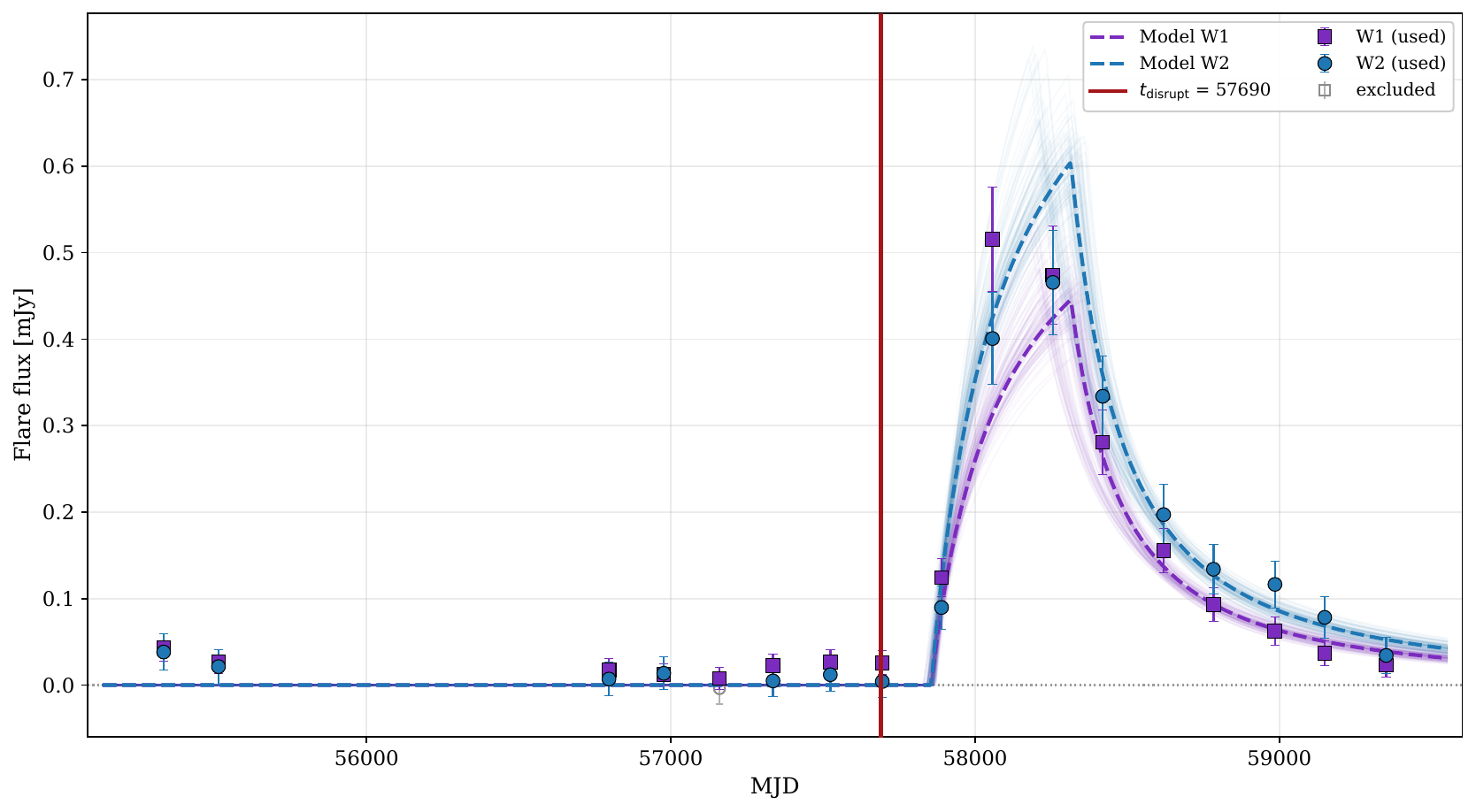}
    \caption{Same as Figure \ref{fig:echomodel_J2244} but for J2150$-$1132 showing the best fit dust echo model.}
    \label{fig:echomodel_J2150}
\end{figure*}

Figure~\ref{fig:echomodel_J2244} shows the fit for J2244$+$0816.  The model reproduces both bands well, giving $R_{\rm dust} = 0.141 \pm 0.015$\,pc, a light-crossing time $2R_{\rm dust}/c = 336$~days, and an implied disruption time $t_{\rm disrupt} = 57370$ (MJD), with $\chi^2/{\rm dof} = 0.62$. The shaded band shows 100 draws from the posterior. Figure~\ref{fig:echomodel_J2150} shows the
same fit for J2150$-$1132, which gives a larger shell, $R_{\rm dust} =
0.193^{+0.012}_{-0.029}$\,pc, a light-crossing time of 460~days, and $t_{\rm disrupt} = 57691$ (MJD).  The fit is noticeably poorer for this source ($\chi^2/{\rm dof} = 1.89$), reflecting a light curve less well described by a single smoothed power-law decline.

Repeating the fit with the temperature shifted by $\pm 1\sigma$ changes $R_{\rm dust}$ by less than 0.002\,pc for either source, and adopting the graybody rather than blackbody amplitude ratio changes it by a comparable amount; the recovered radius is therefore insensitive to the temperature assumption. We caution that $R_{\rm dust}$ enters this model only through the width of the smoothing kernel, and is degenerate with $t_0$: a broader light curve can be produced either by a larger shell or by a slower intrinsic decline. The radii we quote should be read as the shell size required to reproduce the observed smoothing under the assumed decline, not as an independent distance measurement.

\subsection{Tests of the Fallback Power-Law Decline}
\label{sec:fallback_tests}

The spherical-shell model above assumes the canonical $t^{-5/3}$ fallback decline. To test whether this index describes the observed decay, we fit power-law declines of the form $L \propto (t-t_{\rm disrupt})^{-n}$ to the post-peak light curves for $n=5/3$, $9/4$, $3$, and $5$. For J2150$-$1132, which lacks an optical counterpart, the mid-infrared flare is the only available tracer of the decay. For J2244$+$0816, an optical counterpart is present, allowing us to fit the post-peak optical decline after isolating the flare-dominated portion above the host baseline, although the sparse sampling leaves the index only weakly constrained. The optical light curve of J2244$+$0816 is best fit by a power-law index of $n \approx 2.2 \pm 0.5$. While the canonical full-disruption index $n=5/3$ for the complete disruption of a star by a quiescent SMBH \citep{Rees1988, Phinney1989} remains broadly consistent with the data within the uncertainties, a steeper $n=9/4$ decline as expected by the partial disruption of a star by a quiescent SMBH \citep{CoughlinNixon2019} provides a slightly better description of the observed decay. The $n=3$ model is also consistent with the photometry, although it does not improve the fit relative to $n=9/4$, while the steepest $n=5$ decline is disfavored. We therefore cannot conclusively distinguish between a full and a partial disruption nor are our TDE host galaxies inactive, but the observed optical decay shows a slight preference for the steeper $n=9/4$ power-law expected for a partial disruption. The mid-infrared light curves of both J2244$+$0816 and J2150$-$1132 are likewise consistent with power-law indices between $n=5/3$ and $n=3$, and do not discriminate among them given the photometric uncertainties and sparse sampling. We caution that the observed declines trace reprocessed emission from the disk, wind, and dust rather than the intrinsic fallback rate alone.

\subsection{Energetics and the Mass of the Disrupted Star}\label{sec:energetics}

The total energy radiated by each flare is the most direct constraint we have on what powered it, and in particular on how much stellar material was accreted. We therefore integrate the bolometric infrared luminosity over the flare window and use the result to estimate a progenitor mass. For each epoch with a constrained dust temperature, we integrate the best-fit spectrum from 1 to 1000\,\micron\ to obtain a bolometric luminosity, and then integrate $L_{\rm IR}(t)$ over the flare window using the trapezoidal rule \citep[e.g.,][]{2024ApJ...964..117S}, working in rest-frame time throughout. We report results for both the blackbody and graybody models. Table~\ref{tab:modeling_results} collects the modeling and energetics results.

For J2244$+$0816 the flare window spans MJD 57546--59172, or 1454 rest-frame days, giving $\log(E_{\rm rad,BB}/{\rm erg}) = 51.67$ and $\log(E_{\rm rad,GB}/{\rm erg}) = 51.51$, with a peak bolometric luminosity of $\log(L_{\rm peak,BB}/{\rm erg\,s^{-1}}) = 44.18$.  Two-thirds of this energy is contributed by the first two intervals of the light curve, MJD 57546--57910, so the total is dominated by the epochs closest to peak. Integrating the baseline-subtracted optical light curve gives a higher energy, $\log(E_{\rm rad,opt}/{\rm erg}) = 52.1$, as expected if the dust does not completely enshroud the optical flare.

For J2150$-$1132 the window spans MJD 57890--59351, or 1159 rest-frame days, giving $\log(E_{\rm rad,BB}/{\rm erg}) = 51.49$ and $\log(E_{\rm rad,GB}/{\rm erg}) = 51.39$, with $\log(L_{\rm peak,BB}/{\rm erg\,s^{-1}}) = 44.12$.  Because epochs 9 and 10 are excluded on temperature grounds (Section~\ref{sec:temperature}), this integral covers the decay only and omits the rise; retaining those two epochs would raise $\log E_{\rm rad,BB}$ to 51.83. The quoted value is therefore a firm lower bound on the flare energy.

We can turn these energies into a rough progenitor mass.  In a TDE roughly half the disrupted star remains bound to the black hole and is eventually accreted \citep{Rees1988}, so $E_{\rm rad} = \eta M_{\rm bound} c^2$. An accretion efficiency $\eta = 0.1$ implies a disrupted stellar mass of $\gtrsim 0.2\,M_\odot$ for J2244$+$0816 based on the optical flare energy.

This is a lower bound for several reasons: we apply no bolometric correction to the optical flare energy; we make no correction for the dust covering factor, so the mid-infrared energy captures only the reprocessed fraction; our temporal coverage is incomplete, leaving the rise and peak unconstrained; and the efficiency of super-Eddington accretion is itself uncertain and may be well below 0.1 \citep[e.g.,][]{2019ApJ...880...67J}.  Accounting for these, the true disrupted mass is plausibly in the range $0.5$--$1\,M_\odot$.


\begin{deluxetable}{ccc}
\tablecaption{Dust shell modeling and energetics results.}
\label{tab:modeling_results}
\tablehead{\colhead{Property} & \colhead{J2244$+$0816} & \colhead{J2150$-$1132}}
\startdata
Flare window (MJD) & 57546--59172 & 57890--59351 \\
$\langle T_{\rm BB}\rangle$ (K) & $1336 \pm 171$ & $1172 \pm 106$ \\
$\langle T_{\rm GB}\rangle$ (K) & 788 & 785 \\
$\langle R_{\rm BB}\rangle$ (pc) & 0.0368 & 0.0538 \\
$R_{\rm dust}$ (pc) & $0.141 \pm 0.015$ & $0.193^{+0.012}_{-0.029}$ \\
$2R/c$ (days) & 336 & 460 \\
$t_{\rm disrupt}$ (MJD) & 57370 & 57691 \\
$\chi^2/{\rm dof}$ & 0.62 & 1.89 \\
$\log(L_{\rm peak,BB})$ (erg\,s$^{-1}$) & 44.18 & 44.12 \\
$\log(L_{\rm peak,GB})$ (erg\,s$^{-1}$) & 43.98 & 44.00 \\
$\log(E_{\rm rad,BB})$ (erg) & 51.67 & 51.49 \\
$\log(E_{\rm rad,GB})$ (erg) & 51.51 & 51.39 \\
$M_{\rm opt,peak}$ (AB)$^a$ & $-23.4$ & \nodata \\
$\tau_{50}^{\rm opt}$ (rest-frame, days)$^a$ & 77 & \nodata \\
$\nu L_{\nu,\rm opt,peak}$ (erg\,s$^{-1}$)$^a$ & $5.6 \times 10^{44}$ & \nodata \\
\enddata
\tablenotetext{a}{From optical photometry (J2244$+$0816 only).}
\end{deluxetable}

\subsection{Phase Space of Mid-Infrared Transients}
\label{sec:phasespace}

\begin{figure*}[t]
    \centering
    \includegraphics[width=0.85\textwidth]{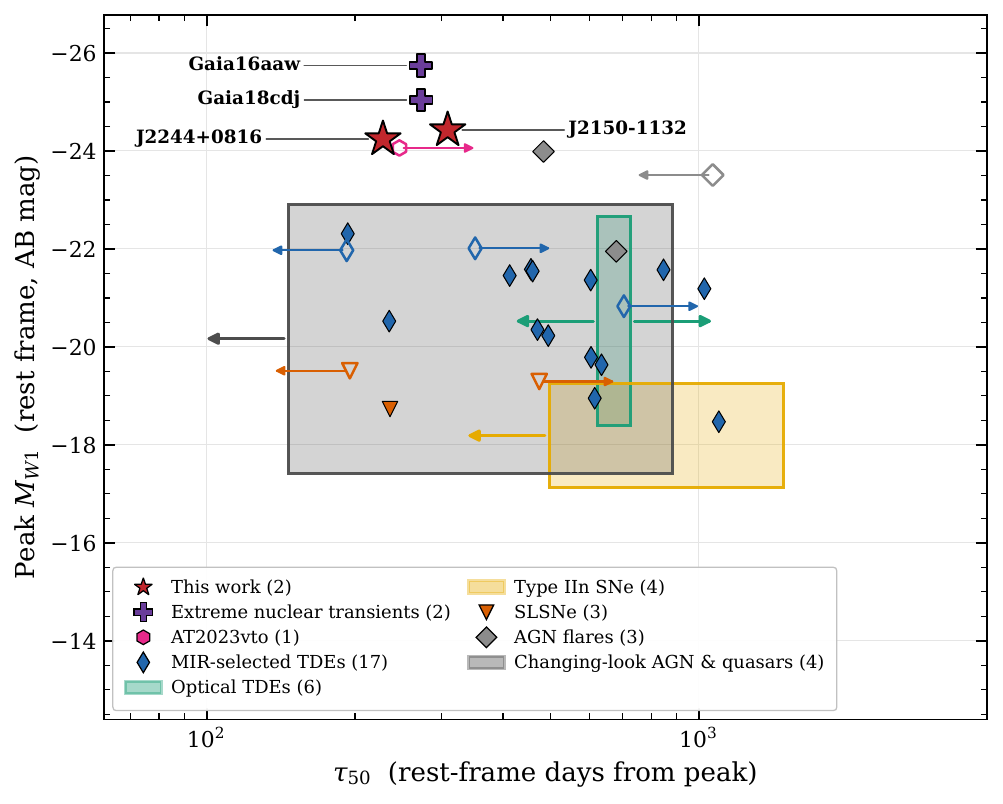}
    \caption{Mid-infrared phase space of luminous transients: peak W1 absolute magnitude ($K$-corrected to rest-frame 3.4~\micron) versus rest-frame $\tau_{50}^{\rm IR}$, the time to fade to half of the peak flux.  All 42 objects are measured homogeneously from unTimely photometry \citep{Meisner2023} after baseline subtraction.  Shaded rectangles span  optically selected TDEs, Type~IIn SNe, and changing-look AGN and quasars, with the horizontal extent set by objects having a measured $\tau_{50}^{\rm IR}$ and the vertical extent by all objects in the class; arrows projecting from a box edge indicate that the class contains objects whose decline is unresolved by the six-month cadence (leftward) or incomplete within the coverage (rightward).  Classes with fewer than three measured objects, i.e., AGN flares, SLSNe, and AT2023vto, are shown as individual points. Mid-infrared-selected TDEs from \citet{masterson2024newpopulationmidinfraredselectedtidal} are shown as blue diamonds. Red stars mark J2244$+$0816 and J2150$-$1132 (this work); purple crosses mark the ENT anchors Gaia16aaw and Gaia18cdj \citep{Hinkle2025}.}
    \label{fig:phase_space}
\end{figure*}

The peak luminosities and decay timescales derived above place both sources in a sparsely populated region of transient phase space in the mid-infrared regime. Figure~\ref{fig:phase_space} shows the peak W1 absolute magnitude ($K$-corrected to rest-frame 3.4~\micron) versus the rest-frame time to fade to half of the peak flux ($\tau_{50}^{\rm IR}$) for J2244$+$0816 and J2150$-$1132, compared to a sample of mid-infrared transients measured in the same way.  This plot mimics a similar analysis by \citet{Hinkle2025} for optical data, though we note that the mid-infrared comparison samples are considerably smaller: WISE is neither deep enough nor rapidly enough cadenced to characterize most supernova classes at these wavelengths.  The extreme nuclear transients (ENTs) Gaia16aaw and Gaia18cdj \citep{Hinkle2025}, along with the luminous helium TDE AT2023vto \citep{HKumar2024}, are plotted individually.

To enable a consistent comparison, 80 sources across a range of flare classes were identified and measured identically: we obtained per-epoch W1 and W2 photometry from the unTimely Catalog \citep{Meisner2023}, which provides time-resolved coadds of the WISE and NEOWISE imaging from 2010 through 2020 at
the nominal six-month cadence.  Detections were matched to each target position within $3.5^{\prime\prime}$, and we required at least ten epochs with a W1 detection. This was satisfied by  57 sources.  Each light curve was baseline-subtracted to isolate the flare: where a published event epoch is available, we average all epochs preceding it by more than 180~days (34 sources).  Otherwise, we adopt the inverse-variance weighted mean of the three faintest epochs (23 sources).
Peak magnitudes were $K$-corrected assuming a blackbody spectrum at the temperature implied by the baseline-subtracted W1/W2 flux ratio at peak (38 sources), or, where that colour is unconstrained, at a fiducial 1200~K (19 sources); the correction is smaller than 0.1~mag for the $z < 0.05$ objects that comprise most of the comparison sample.  Objects retaining less than 20\% of their observed peak flux after subtraction were excluded, since for these the measurement is a small difference between two similar numbers and the host-subtraction systematic dominates.  Objects whose post-peak decline is unresolved by the six-month cadence are shown as upper limits on $\tau_{50}^{\rm IR}$, and objects whose decline has not completed within the available coverage are shown as lower limits.  In total, 42 of the 80 sources searched satisfy all these criteria and appear in Figure~\ref{fig:phase_space}.

The comparison sample comprises the mid-infrared-selected TDEs of \citet{masterson2024newpopulationmidinfraredselectedtidal}; optically selected TDEs with reported dust echoes \citep[from][]{Gezari2012, arcavi, Holoien2016a, Holoien2016b, Blagorodnova2017, mattila, nicholl, Gezari2020ATel, stein, Wevers2021, Onori2022, Cao2024}, including ASASSN-15lh \citep{Dong2016, Leloudas2016}; Type~IIn SNe \citep[from][]{Ofek2007, Miller2010, Stoll2011, Fox2011, Fox2013, Andrews2017, BKumar2019, Szalai2019}; SLSNe, drawn primarily from the WISE study of \citet{2022MNRAS.513.4057S} \citep[see also][]{Nicholl2013, Nicholl2016, Anderson2018}; and AGN flares and changing-look AGN \citep[from][]{McElroy2016, Kankare2017, Yang2018, Arcavi2019, Trakhtenbrot2019, Graham2020, Malyali2021, Oknyansky2021, Yu2022}.

We also searched for mid-infrared counterparts to three fast blue optical transients \citep{Perley2019, Coppejans2020, Perley2021} and three thermal supernovae \citep{Nugent2011, Fossey2014, Jencson2023}, but recovered no unTimely detections at any of these positions: these events are either too faint in the mid-infrared or evolve too rapidly to be captured within a single WISE sky pass, and they are therefore absent from the figure.

Our flaring sources lie at $M_{\rm W1} = -24.2$ (J2244$+$0816) and $-24.4$ (J2150$-$1132), roughly two magnitudes brighter than the regions occupied by the supernova classes measurable here, including superluminous supernovae. Our events instead fall within the regime occupied by extreme nuclear transients (ENTs) and other luminous accretion-powered transients.  We caution that the wider range in $\tau_{50}^{\rm IR}$ spanned by the \citet{masterson2024newpopulationmidinfraredselectedtidal} sample relative to the optically selected TDEs reflects their selection on mid-infrared flare strength rather than an intrinsic difference between the two populations: optically
selected TDEs enter our sample only when their dust echo is luminous enough to survive the subtraction described above.  We further note that for a small number of objects, including J2150$-$1132, the peak-epoch color implies a temperature above the graphite sublimation limit, indicating that the emission at peak is not purely reprocessed dust; the corresponding $K$-corrections are correspondingly uncertain, though they remain small at these redshifts. The vertical separation, however, is robust: both sources remain more luminous than every comparison class whether the baseline is taken from pre-flare epochs, from the faintest epochs, or omitted entirely.

\section{Discussion}\label{sec:discussion}

The two sources presented in this work show luminous mid-infrared flares consistent with dust echoes of energetic nuclear transients (\S~6.1), most likely from a TDE (\S~6.2). For J2244$+$0816 (\S~6.3), we argue that the combination of a fast optical flare, a multi-year mid-infrared echo, and post-flare AGN dimming is best explained by a TDE---potentially a micro-TDE by an embedded stellar-mass black hole within the AGN accretion disk. For J2150$-$1132 (\S~6.4), the changing-look spectroscopic evolution and absence of an optical counterpart point to a dust-obscured TDE.

\subsection{Comparison with Other Mid-Infrared TDEs}
\label{sec:comparison_discussion}

Our dust radii of 0.14 and 0.19~pc fit well within the 0.05--0.46~pc range from \citet{masterson2024newpopulationmidinfraredselectedtidal}, who found a mean of $0.23 \pm 0.11$~pc. Our light-crossing times of 336 and 460~days are also consistent with their 60--550~day range. However, our peak bolometric luminosities of $\log(L_{\mathrm{peak,BB}} / \mathrm{erg}$ $\mathrm{s^{-1}}$) = 44.18 and 44.12 are brighter than the peak W2 luminosities of their distribution, which spans values of 42.1 to 43.4, perhaps due to the different geometry or propertiest of circumnuclear dust between active and inactive galaxies. 

The flare-averaged blackbody temperatures of $1336 \pm 171$~K and $1172 \pm 106$~K, with the hottest well-constrained epochs approaching the sublimation temperature, match what we expect for dust heated close to sublimation \citep{Lu2016, vanVelzen2016}. \citet{Jiang2021} showed that mid-infrared dust echoes are common even among optically discovered TDEs, and \citep{vanVelzen2024} applied a related echo selection to 3142 ZTF nuclear transients, identifying 63 accretion flares with candidate dust echoes and unifying classical TDEs with extreme AGN flares under a single infrared signature. Our mid-infrared-selected candidates extend this picture to the obscured regime: we report an event that is undetected in the optical but bright in the mid-infrared, selected on the basis of WISE variability. The higher peak luminosities we measure potentially reflects the fact that optical selection misses the most heavily obscured events, which would be able reprocess a larger fraction of the flare energy into the infrared.  In addition, our selection identifies events in active galaxies while \citet{masterson2024newpopulationmidinfraredselectedtidal} studied events in quiescent galaxies.  Active galaxies have accretion disks and presumably different dusty enshrouding regions which could both augment the mid-infared flare energy.


\subsection{Non-TDE Scenarios}
\label{sec:sn_rejection}

We next discuss potential physical scenarios to explain the two reported flares, starting with non-TDE events and considering TDEs in the following sections.

\subsubsection{Microlensing}
Gravitational microlensing by a foreground star produces symmetric, achromatic flares with characteristic timescales of months to $\sim$2~yr \citep{Graham2017}. The mid-infrared flares in both candidates span $\gtrsim$5~yr and show significant W1$-$W2 color evolution, ruling out this scenario.

\subsubsection{Normal Supernovae}
Core-collapse and Type~Ia supernovae peak at $\sim$$10^{43}$~erg~s$^{-1}$ in the optical and fade within months \citep{Richardson2014}. The peak mid-infrared luminosities of both candidates exceed $10^{44}$~erg~s$^{-1}$, and the total radiated energies, $E_{\rm rad} \approx 10^{51.5}$--$10^{51.7}\, {\rm erg}$,  exceed the total energy output of any normal supernova by two to three orders of magnitude. In the mid-infrared specifically, the brightest known supernova (SN~2010jl, a Type~IIn) peaks at $\sim$$10^{42}$~erg~s$^{-1}$ in the Spitzer bands \citep{Szalai2019}, 100$\times$ fainter than our sources, ruling out a normal supernova as a plausible scenario.

\subsubsection{Superluminous Supernovae}
SLSNe can reach peak luminosities of $\sim$$10^{44}$--$10^{45}$~erg~s$^{-1}$ in the optical \citep{GalYam2012}, overlapping with our observed flare luminosities. However, several properties of J2244$+$0816 and J2150$-$1132 are inconsistent with the SLSN interpretation in terms of their energetics and luminosities. 

SLSNe have not been observed to produce multi-year, parsec-scale mid-infrared dust echoes. \citet{2022MNRAS.513.4057S} present WISE W1 and W2 light curves of 10 SLSNe at $z < 0.12$; for the two objects with the most luminous reprocessed dust emission (PS15br and SN~2017ens), their modeled dust SEDs reach mid-infrared luminosities of only $10^{41.5}$--$10^{42.5}$~erg~s$^{-1}$, fading within $\sim$1--2~years. The peak mid-infrared luminosities of J2244$+$0816 ($10^{44.2}$~erg~s$^{-1}$) and J2150$-$1132 ($10^{44.1}$~erg~s$^{-1}$) exceed the brightest SLSN in the \citet{2022MNRAS.513.4057S} sample by more than 1.5 orders of magnitude, and both mid-infrared flares persist for $\gtrsim$3--4~years. In the Spitzer bands, even the most luminous Type~IIn supernova (SN~2010jl) peaks at $\sim$$10^{42}$~erg~s$^{-1}$ \citep{Szalai2019}, more than two orders of magnitude below our sources. The total mid-infrared radiated energies, $\log(E_{\rm rad}/{\rm erg}) = 51.67$ (J2244$+$0816) and $51.49$ (J2150$-$1132), exceed the \emph{total} radiated energy of SLSNe ($\sim$$10^{51}$~erg; \citealt{GalYam2012}) by factors of $\sim$5 and $\sim$3 respectively, despite representing only the dust-reprocessed component of each flare and, for J2150$-$1132, only the decay phase. For comparison, the only SLSN-like event with comparable energetics is ASASSN-15lh \citep{Dong2016}, which has since been reclassified as a TDE \citep{Leloudas2016}. Together, these arguments rule out SLSNe as the origin of the two flares, and the preceding sections rule out a variety of non-TDE origins for the flares.  In the following, we discuss potential TDE origins for the two flares.


\subsection{J2244 $+$ 0816} 
\label{sec:microtde}

The multi-wavelength properties of J2244$+$0816 present a distinctive picture: a fast optical flare ($\tau_{50}^{\rm opt} = 77$~rest-frame days, $M_{\rm opt,peak} = -23.4$) followed by a slow, multi-year mid-infrared dust echo ($R_{\rm dust} = 0.14$~pc, $\langle T_{\rm BB}\rangle \sim 1340$~K). In addition, the immediate post-flare mid-infrared luminosity is slightly depressed with respect to the pre-flare and late-time luminosity and we find no strong evidence for broad-line spectral  changes across the two epochs of spectroscopy. We consider two scenarios involving tidal disruptions in the AGN environment.

\subsubsection{Scenario 1: TDE by an SMBH} 


In this scenario, a star from the nuclear star cluster is disrupted by the central SMBH, similar to the majority of TDEs in the literature \citep[e.g.,][]{Gezari2021}. The mid-infrared echo arises from dust reprocessing at the torus scale; this corresponds to the standard interpretation for mid-infrared-selected TDEs \citep{vanVelzen2016, masterson2024newpopulationmidinfraredselectedtidal}. The biggest challenge to this picture is that TDEs in the literature are generally, by design, restricted to quiescent SMBHs and the observed optical decay of J2244+0816 is much faster than the expected intrinsic $t^{-5/3}$ fallback rate for such events. So, while the observed phenomenology of the J2244+0816 flare does not fully comport to the expectations of a classical TDE, they are not expected to for a TDE in an active galaxy.

The pre-existing accretion disk potentially provides a natural explanation for the more unusual features of this flare.  Based on hydrodynamical simulations of TDEs in AGN discs, \citet{Ryu2024} show that the observed properties of such a TDE will differ substantially from a `naked' TDE around a quiescent SMBH, depending on the disc density and the orientation of the stellar orbit with respect to the accretion disk (i.e., prograde vs. retrograde orbits of the disrupted star with respect to the accretion disk). \citet{McKernan2022} find that retrograde AGN tidal disruptions yield brighter but shorter-lived flares as in-plane ejecta collide with the inner accretion disc, offering a route to the rapid decay we observe. They also predict that the collision with the inner accretion disk will deplete the inner disk, plausibly accounting for the AGN dimming seen after the flare \citep[see also][]{Chan2019}. More broadly, \citet{Wang2024CL} argue that AGN TDEs may constitute a fraction of observed changing-look AGN, connecting events such as inferred here to longer-term AGN state changes. 

Finally, we note that the empirical $K$-band $R_{\rm dust}–L$ relation from \citet{Koshida2014} predicts that the inner edge of the dust torus of a $\sim 10^{44}\, {\rm erg}\, {\rm s}^{-1}$ AGN should be at distance of $\sim 0.1$~pc from the SMBH, implying a lag of $\sim 120$~days between a brightening event close to the central SMBH and its infrared dust echo.  Based on Fig.~\ref{fig:Plot_J2244_0816}, we measure a plausibly consistent lag of $\leq 105$~days between the optical and mid-infrared flare peaks, where the limit is due to the sparse, $\sim 180$~day sampling of the WISE data.  Had the optical and mid-infrared peaks been simultaneous, we would not know; indeed, the WISE data are, in principle, consistent with a mid-infrared peak preceding the optical peak, though that is physically unlikely.  In short, the measured time lag is consistent with a dust echo of an event close to the SMBH, but is also consistent with the dust echo of an event farther out in the accretion disk, which would yield a shorter lag.

\subsubsection{Scenario 2: $\mu$TDE by a Stellar-Mass Black Hole }
 
An alternative explanation is that the flare comes from a ``micro-TDE'', the tidal disruption of a star by a stellar-mass black hole (sBH) embedded in the AGN accretion disk. As discussed in \citet{Yang2022}, in an AGN disk, stars and sBHs migrate through the gas and can become trapped at the same locations. Three-body interactions can then scatter a star within the tidal radius of an sBH, leading to a tidal disruption. \citet{Yang2022} predict a micro-TDE rate of $\sim$ 170~Gpc$^{-3}$~yr$^{-1}$, which is approximately half the volumetric rate for optical TDEs inferred by \citet{Yao2023}.  However, no micro-TDEs have been confirmed to date.
 
Several features of J2244$+$0816 are moderately consistent with this scenario:
 
\emph{Fast optical rise and steep decay.} A micro-TDE evolves quickly. The disrupting object is an sBH, potentially reaching into the intermediate mass black hole (IMBH) regime, so the fallback timescale is short and the late-time fallback is steeper than the $\dot{M} \propto t^{-5/3}$ scaling of a full SMBH disruption. In this regime, the surviving stellar core pulls on the debris stream and pushes the fallback away from the canonical $t^{-5/3}$ expectation \citep{Wang2021}. Low-eccentricity orbits in the disk favor partial disruptions \citep{Wang2021}, whose late-time fallback approaches $\dot{M} \propto t^{-9/4}$ independent of the core mass \citep{CoughlinNixon2019}. Hydrodynamical simulations of partial stellar-mass TDEs find slopes of $\approx t^{-2.1}$ to $t^{-1.71}$ \citep{Wang2021} and heavily depend on the mass of the black hole, with lower mass sBHs having steeper fallback rates \citep{CoughlinNixon2019, Wang2021}. A steep decline is therefore expected for a low-mass disruptor and is not unusual. As mentioned in Section \ref{sec:fallback_tests}, the power-law decline of the J2244+0816 optical data favor steep, $t^{-9/4}$ and $t^{-3}$ power-law declines, suggestive of a partial disruption by a lower mass black hole.

\emph{AGN-disk environment.} The AGN supplies the dense medium the model needs. The disk gas around the disruption site delays and stretches the light curve \citep{Kremer2019, Yang2022}. Whether the signal escapes, and how strongly, depends on the disk mass and the location of the event in the disk \citep{Perna2021a}. Micro-TDEs are expected at $\sim$$10^{-4}$--$10^{-2}$~pc from the SMBH, where the disk is ionized and opaque. There the sBH can open a cavity that lowers the opacity and lets the transient emerge \citep{Yang2022}. This makes it possible for a stellar-mass disruption to produce the observed optical flare and later the extended, infrared-bright signal.
 
The main problem for the micro-TDE interpretation is the energetics. \citet{Kremer2023} predict peak bolometric luminosities of $\sim$$10^{41}$--$10^{44}$~erg~s$^{-1}$, set mainly by the wind-to-accretion partition (their $s$ parameter). The mass of the disrupted star and the penetration factor matter only at a factor of a few level. The peak mid-infrared bolometric luminosity of J2244$+$0816, $\log(L_{\rm peak,BB}/{\rm erg\,s^{-1}}) = 44.18$, already sits at the top of this range and the baseline-subtracted peak optical luminosity, $\log(\nu L_{\rm \nu,opt}/{\rm erg\,s^{-1}}) = 44.75$, exceeds it by a factor of four. The total mid-infrared radiated energy, $\log(E_{\rm rad}/{\rm erg}) = 51.67$, is a lower bound, since it assumes a spherical dust shell absorbing and reprocessing the entire UV/optical event. Matching these values requires one of a few options. The embedded black hole could be relatively massive ($M_{\rm BH}\sim50$--$100\,M_\odot$). This is plausible in AGN disks, where sBHs grow by gas accretion \citep{McKernan2012} and mergers \citep{2025ApJ...990..217M}. In this regime, \citet{Kremer2023} find luminosities that approach the brightest fast blue optical transients such as AT2018cow. The accretion could also be sustained and super-Eddington. Or, most plausibly, the SMBH itself could contribute as the micro-TDE perturbs and destabilizes the inner disk, potentially increasing the SMBH accretion rate for a period of time. In that case the observed luminosity is a sum of the micro-TDE and the SMBH's response.

\subsection{J2150-1132} 
\label{sec:j2150_discussion}
J2150$-$1132 presents a different observational picture. The absence of detectable optical variability combined with the strong mid-infrared flare and the changing-look spectroscopic evolution points to a heavily dust-obscured TDE. The DESI spectrum taken $\sim$3~years after the mid-infrared peak shows broadened H$\alpha$ emission (type~1.9 AGN), while the NGPS spectrum taken $\sim$4~years after the DESI spectrum shows only narrow emission lines. This ``turn-off'' of the broad-line region is characteristic of changing-look AGN \citep[e.g.,][]{LaMassa2015, MacLeod2016, Graham2017, Graham2020, Ricci2020} and fits naturally with TDE-driven changing-look events, in which the tidal disruption temporarily powers the accretion disk and illuminates the broad-line region, which then fades as the fallback rate drops \citep{Merloni2015, Trakhtenbrot2019, Wang2024CL}. The total radiated energy of $\log(E_{\rm rad}/{\rm erg}) = 51.49$, which covers only the decay phase of the flare, is consistent with a canonical SMBH TDE.

The detection of a coherent mid-infrared flare itself constrains the density of the inner accretion disc. Based on hydrodynamical simulations of a TDE in an AGN, \citet{Ryu2024} show that very low disc densities will produce a TDE indistinguishable from a ``naked'' TDE in a quiescent galaxy, while for very high disc densities, the debris mixes into the disc and no flare is produced. The presence of a mid-infrared flare in J2150$-$1132 therefore constrains the disc density to an intermediate value, while the accompanying changing-look behavior implies the event was able to perturb the inner region and drive a state change in the inner disc.  
The high luminosity of the flare and subsequent turn-off of the broad-line region and depressed post-flare mid-infrared luminosity are all consistent with the predictions for a retrograde AGN-TDE in which in-plane debris from the tidal disruption collides with the inner accretion disc to drive an over-luminous flare followed by a lower AGN continuum state as the disc refills on the viscous timescale \citep{McKernan2022}.  In contrast and unlike the flare from J2150-1132, \citet{McKernan2022} predict that a prograde AGN-TDE will add angular momentum to inner disc gas and initially look like a regular TDE but then will be followed by an AGN high state. Hydrodynamic simulations support this orientation dependence, with retrograde and high-inclination encounters producing brighter, shorter-lived flares than prograde encounters \citep{2025ApJ...993..244Z}.

The combination of a luminous mid-infrared flare ($L_{\rm peak,BB} \approx 10^{44.1}\, {\rm erg\,s^{-1}}$), no optical flare counterpart (suggesting heavy dust obscuration), and changing-look spectroscopy caused by the inner accretion disk state changes make J2150$-$1132 a strong candidate for an obscured TDE in an AGN host, extending the mid-infrared TDE population in quiescent galaxies from \citet{masterson2024newpopulationmidinfraredselectedtidal} to TDEs in active galaxies  from the WISE R90 AGN catalog.  Current data cannot firmly establish whether the mid-infrared flare is the dust echo of a micro-TDE or an SMBH TDE, though, in the latter case, the data prefer a retrograde TDE over a prograde TDE.


\section{Conclusions}\label{sec:conclusions}

We investigated extreme mid-infrared variability in WISE-selected AGN and found two strong TDE candidates: J2244$+$0816 at $z = 0.1185$ and J2150$-$1132 at $z = 0.2605$. Both show large mid-infrared flares consistent with dust echoes of events closer to the AGN central engine.  Modeling the  flares with a thin, spherical dust model implies dust radii of 0.14 and 0.19\,pc, respectively, and total radiated energies of $\log(E_{\rm rad}/{\rm erg}) = 51.67$ and 51.49, respectively, for the two events.  These exceed the total radiated energy of superluminous supernovae by factors of a few, and that of normal supernovae by more than two orders of magnitude.

For J2244$+$0816, we report an optical counterpart flare to the mid-infrared event that peaks at $M_{\rm opt,peak} = -23.4$ with a rest-frame decay timescale of 77~days to reach 50\% of its peak optical flux value.  The mid-infrared decay is consistent with the canonical $t^{-5/3}$ fallback decline, although a partial disruption ($t^{-9/4}$) cannot be excluded. The optical flare continues to decay to a level $\sim 0.6$~mag below the pre-flare baseline, before eventually recovering.  A similar post-flare depression of the mid-infrared luminosity is also detected.  We argue that these properties are best explained by a TDE in an AGN.  If the flare is due to a TDE by an SMBH, the post-flare depression implies that the star's orbit was retrograde with respect to the accretion disk rotation, and the debris from the depression affected the inner accretion disk.  We also note that the event could be due to a micro-TDE by a stellar- or intermediate-mass black hole embedded in the AGN accretion disk. This interpretation is supported by recent theoretical predictions of micro-TDEs in AGN disk environments \citep[e.g.,][]{Ryu2024}.

For J2150$-$1132, which is a type-2, or obscured, AGN based on optical spectroscopy, we detect no optical counterpart to the mid-infrared flare. This indicates the first detection of the expected population of TDEs in obscured, or type-2 AGN. Again, the post-flare mid-infrared luminosity is depressed relative to the pre-flare level, suggestive of a retrograde, rather than prograde, TDE. Optical spectroscopic data for J2150$-$1132 taken three years after the mid-infrared peak shows broad H$\alpha$ emission, which disappeared in subsequent spectroscopy.  This is consistent with underlying state changes or a changing-look AGN caused by the collision between incoming tidal disruption debris and the inner accretion disk, as predicted by \citet{McKernan2022}.

These results demonstrate the potential of the WISE mid-infrared survey to find the full census of TDEs in AGN, including events in obscured, or type-2 AGN that are missed by X-ray, UV, and optical surveys.  The mid-infrared emission provides a powerful dust echo bolometer of the inner central engine, providing key information about the event including its energetics, the dust geometry, and even the nature of the tidal disruption given the distinct predictions of prograde vs retrograde TDEs in AGN. This work motivates future modeling work on such events, including micro-TDEs where the disrupting black hole resides within the accretion disk, as well as future observations with improved cadence.

\bigskip
\section*{Acknowledgments}

We thank Maya Nunez and Luis E. Olague Rodriguez for their assistance in obtaining the Palomar observations presented in this work.

The work of D.S. was carried out at the Jet Propulsion Laboratory, California Institute of Technology, under a contract with the National Aeronautics and Space Administration (80NM0018D0004). M.J.G. acknowledges support by the US National Science Foundation (NSF) through grant AST-2108402. B.M. and K.E.S.F. are supported by NSF AST-1831415, NSF AST-2206096, and Simons Foundation Grant 533845. We acknowledge the support of the National Aeronautics and Space Administration through ADAP grant number 80NSSC24K0663.

This publication makes use of data products from the Wide-field Infrared Survey Explorer, which is a joint project of the University of California, Los Angeles, and the Jet Propulsion Laboratory/California Institute of Technology, funded by the National Aeronautics and Space Administration. 

Portions of the manuscript text were edited for grammar, clarity, and readability with assistance from Claude (Anthropic;
\citealt{anthropic_claude_2026}). This tool was not used to generate scientific content, perform analysis, interpret results, draw conclusions, or produce citations. All results, interpretations, and conclusions presented here are the original work of the author, who reviewed the final text and take full responsibility for its accuracy.

\section*{Author Contributions}
S.T.A.U. led the analysis and wrote the manuscript. D.S. and D.H.M. supervised the project and provided detailed feedback on the manuscript. K.D. and C.P. provided the WISE R90 catalog used for sample selection. J.A.A.B. assisted with the acquisition of the Palomar spectra. M.J.G. carried out the optical counterpart confirmation, and provided the optical data for this work. M.B. reduced the XMM-Newton data and contributed the text describing the X-ray follow-up and its interpretation. B.M. and K.E.S.F. contributed the theoretical interpretation of both the sources.


\bibliography{main}{}
\bibliographystyle{aasjournal}

\end{document}